\documentclass[12pt]{article}
\pdfoutput=1
\usepackage{diagbox}
\usepackage{subcaption}
\usepackage{amsmath}

\usepackage{amssymb}
\usepackage{bm}
\usepackage{dsfont}
\usepackage{cite}
\usepackage{textcomp}
\usepackage{calc}
\usepackage{geometry}
\usepackage{bbm}
\usepackage[utf8]{inputenc}
\usepackage[normalem]{ulem}
\usepackage[english]{babel}
\usepackage{mathtools}
\usepackage{graphicx}
\usepackage[affil-it]{authblk}
\usepackage{color}
\usepackage{verbatim}
\usepackage{slashed}
\usepackage{cancel}
\usepackage{braket}
\usepackage{multicol}
\usepackage{tcolorbox}
\tcbuselibrary{theorems}
\usepackage{enumitem}
\usepackage{float}
\usepackage{soul}
\usepackage[toc,page]{appendix}
\usepackage{hyperref}
\hypersetup{colorlinks=true,urlcolor=blue,linkcolor=red,citecolor=green!60!black}

\newcommand{\be}{\begin{equation}}
\newcommand{\ee}{\end{equation}}
\newcommand{\gss}{g_{\star s}}
\newcommand{\gs}{g_\star}
\newcommand{\TRH}{T_{\rm RH}}
\newcommand{\rp}{\rho_\phi}
\newcommand{\rR}{\rho_r}
\newcommand{\LambdaQCD}{\Lambda_{\text{QCD}}}
\newcommand{\Gphi}{\Gamma_\phi}
\newcommand{\ma}{m_a}
\newcommand{\maz}{m_{a,0}}
\newcommand{\Tl}{T_\Lambda}
\newcommand{\TQCD}{T_{\rm QCD}}
\newcommand{\RRH}{R_{\rm RH}}
\newcommand{\kRH}{k_{\rm RH}}
\newcommand{\Tosc}{T_\text{osc}}
\newcommand{\Rosc}{R_\text{osc}}
\newcommand{\kosc}{k_\text{osc}}
\newcommand{\kini}{k_\text{ini}}
\newcommand{\Tbbn}{T_\text{BBN}}
\newcommand{\thini}{\theta_\text{ini}}
\newcommand{\Rini}{R_\text{ini}}

\hypersetup{linktocpage=true}

\begin{document}
\clearpage\thispagestyle{empty}

\begin{center}
    \vspace*{0.5cm}
    {\LARGE{\textbf{From Rags to Jeans:\\Axion Miniclusters\\from Early matter domination}}}\\
    \date{}
    \vspace*{1.2cm}

    {\large{Ariel Angulo$^{1}$, Paola Arias$^{2}$,\\
    Nicolás Bernal$^{3}$ and Javier Redondo$^{4}$}}\\[3mm]
    \it{$^1$Departamento de Física, Universidad de Santiago de Chile,\\
    Av. Víctor Jara 3493, Estación Central, Santiago, Chile} \\
    \it{$^2$Departamento de Física, Universidad Técnica Federico Santa María, \\Casilla 110-V, Avda. España 1680, Valparaíso, Chile} \\
    \it{$^3$New York University Abu Dhabi\\
    PO Box 129188, Saadiyat Island, Abu Dhabi, United Arab Emirates}\\
    \it{$^4$CAPA \& Departamento de F\'isica Te\'orica, Universidad de Zaragoza, 50009 Zaragoza, Spain}
\end{center}

\begin{abstract}
In an early matter-dominated era, density and temperature inhomogeneities of the radiation bath grow more efficiently than in the standard radiation-dominated history. If the axion mass depends on temperature, these inhomogeneities induce spatial fluctuations of the axion mass, providing a new source term for axion density perturbations. We show that this mechanism is most efficient when the reheating temperature lies just below the mass-saturation scale $T_\Lambda$, and can drive axion overdensities to order unity by matter--radiation equality. For the QCD axion saturating the observed dark matter abundance, the nonlinear spectrum at equality exhibits two characteristic regions: one associated with the gravitational enhancement already present in moduli-driven cosmologies, and another produced by the temperature dependence of the axion mass. We estimate the resulting minicluster masses and discuss the possible formation of axion miniclusters and axion-star substructure.
\end{abstract}

\newpage
\tableofcontents
\newpage

\section{Introduction}
\label{sec:introduction}
The nature of dark matter (DM) remains one of the central open questions in fundamental physics. Among the many candidates proposed over the decades, bosonic pseudoscalar particles --- axions and axion-like particles (ALPs) --- occupy a privileged position~\cite{Preskill:1982cy, Stecker:1982ws, Dine:1982ah, Abbott:1982af}: the QCD axion~\cite{Weinberg:1977ma, Wilczek:1977pj} arises as the pseudo-Nambu--Goldstone boson of the Peccei--Quinn (PQ) symmetry~\cite{Peccei:1977hh, Peccei:1977np, Peccei:1977ur}, elegantly solving the strong-CP problem, while string compactifications and other ultraviolet completions of the Standard Model (SM) generically predict a landscape of additional pseudoscalar fields with axion-like couplings, the so-called axiverse~\cite{Arvanitaki:2009fg}. Throughout this work, we use the term ``axion'' to refer to both the QCD axion and ALPs, unless a distinction is explicitly needed.

A defining feature of axion DM is its extreme lightness: with masses typically well below the eV scale, axions are typically not produced thermally in sufficient abundance and instead rely on non-thermal mechanisms. The best known among these is the vacuum misalignment mechanism~\cite{Dine:1982ah, Arias:2012az}, in which the axion field is initially frozen at an arbitrary value by Hubble friction and, once its mass becomes comparable to the expansion rate, begins coherent oscillations around the minimum of its potential. The resulting condensate is highly monochromatic --- oscillating at a frequency set by the axion mass --- and behaves as cold, pressureless matter, providing an excellent DM candidate. Several variants and extensions of this mechanism have been suggested, including kinetic misalignment~\cite{Co:2019jts, Barman:2021rdr}, acoustic misalignment~\cite{Bodas:2025eca}, frictional misalignment~\cite{Papageorgiou:2022prc}, and trapped misalignment~\cite{DiLuzio:2021gos, Nakagawa:2020zjr, Kitajima:2023pby}, all of which exploit different dynamical histories for the axion field prior to the onset of oscillations.

A crucial ingredient in all misalignment-based production scenarios is the possible temperature dependence of the axion mass. For the QCD axion, this dependence is well established: the mass is generated by the topological susceptibility of QCD, $\chi(T)$, which is strongly suppressed at temperatures $T$ above the QCD crossover, $T \gtrsim \LambdaQCD \simeq 150$~MeV, and approaches its zero-temperature value only once non-perturbative effects fully develop. The steep power-law behavior $\ma(T) \propto T^{-b}$ at high temperatures --- with $b \simeq 4$ from lattice simulations --- profoundly affects the onset of oscillations, the relic abundance, and the subsequent cosmological evolution of the axion condensate.

Importantly, temperature-dependent masses are not exclusive to the QCD axion. ALPs arising from the breaking of an approximate global $U(1)$ symmetry can acquire a mass through non-perturbative, instanton-like effects in a hidden confining sector, in complete analogy with the QCD mechanism. If the hidden sector has its own confinement scale $\Lambda$, the ALP mass will exhibit a similar thermal suppression for $T \gtrsim \Lambda$, with the power-law index $b$ and the scale $\Tl$ determined by the gauge group and matter content of the hidden sector rather than by QCD. This observation considerably broadens the phenomenological relevance of a temperature-dependent axion mass: the framework developed in this work applies to any pseudoscalar whose mass turns on as the Universe cools through a confining phase transition, whether visible or hidden.

The temperature dependence of the mass has far-reaching consequences that go beyond the background evolution of the axion field. At the level of cosmological perturbations, a time-varying mass introduces qualitatively new effects. Fluctuations in the radiation temperature directly translate into fluctuations of the axion mass, $\delta \ma \propto (d\ma / dT)\,\delta T$, coupling the axion perturbation equation to the radiation overdensity. This coupling acts as an additional source term for axion density perturbations --- absent in the constant-mass case --- that can significantly enhance the growth of small-scale overdensities~\cite{Sikivie:2021trt, Kitajima:2021inh, Ayad:2025awu, Allali:2025pja}. As we show in this work, the efficiency of this mechanism depends sensitively on the interplay between the temperature scale $\Tl$ at which the mass saturates, the reheating temperature $\TRH$, and the comoving wavenumber of the perturbation mode. A powerful observable consequence of these effects is the imprint left on the matter power spectrum and on the possible formation of compact objects such as axion miniclusters~\cite{Kolb:1993zz, Hardy:2016mns, Visinelli:2018wza, Blinov:2019jqc}.
 
The relic abundance and perturbation spectrum of the axion DM are also sensitive to the expansion history of the Universe prior to Big Bang Nucleosynthesis (BBN)~\cite{Giudice:2000ex, Visinelli:2009kt, Arias:2021rer}. Recently, it has been shown that adiabatic fluctuations cannot grow enough to form dense miniclusters during radiation domination~\cite{Sikivie:2021trt, Kitajima:2021inh}, despite the additional temperature-dependent source. Naturally, the situation can be dramatically different in early cosmologies where adiabatic fluctuations are larger at the relevant scales, for instance driven by a different cosmological expansion. Motivated by these findings, in this paper we consider an early matter-dominated (EMD) era, as realized, for example, in low-temperature reheating (LTR) scenarios. During this matter domination, density and temperature fluctuations grow larger than they would during radiation domination, and, as we shall see, they can lead to $\mathcal{O}(1)$ axion-DM fluctuations and, subsequently, to axion miniclusters and axion stars. The EMD era is driven by a long-lived scalar field $\phi$ that modifies both the Hubble expansion rate and the SM temperature evolution, relative to the standard radiation-dominated picture. In minimal scenarios $\phi$ could be identified with the inflaton during cosmic reheating~\cite{Dolgov:1989us, Traschen:1990sw, Kofman:1994rk, Kofman:1997yn, Allahverdi:2020bys, Batell:2024dsi, Barman:2025lvk}, but could also be a moduli field from string compactification~\cite{Ellis:1986zt, Banks:1993en, Kane:2015jia, Cicoli:2023opf}.

In this paper, we study the growth of axion perturbations during an EMD era with low reheating temperature, paying special attention to the role of a temperature-dependent axion mass. We work within the framework of the standard (vacuum) misalignment mechanism~\cite{Dine:1982ah, Arias:2012az} and consider a general parameterization of the mass, $\ma(T) = \maz\,(\Tl / T)^b$ for $T > \Tl$, which encompasses  ALPs with hidden-sector confinement at arbitrary scales. For the QCD axion, we use the topological susceptibility obtained from lattice results~\cite{Borsanyi:2016ksw}. Rather than adopting a fluid description for the axion --- which becomes technically cumbersome when the mass itself carries perturbations --- we solve the perturbed Klein--Gordon equation directly, using as input the background and perturbed solutions for the decaying scalar field and radiation obtained from the coupled Boltzmann--Einstein system. This approach allows us to capture the full dynamics of the axion field through the onset of oscillations and across the reheating transition, including the new source term induced by the temperature-dependent mass.

The paper is organized as follows. Section~\ref {sec:EMD} describes the EMD phase and the resulting modifications to the cosmological background. In Section~\ref{sec:background}, we review the background solutions for the axion field, including approximate analytical expressions for both constant and temperature-dependent masses. In Section~\ref{sec:perturbations}, we derive the perturbed equations for the axion field in the presence of a temperature-dependent mass and discuss the structure of the source terms. Section~\ref{sec:perturbations} also presents our numerical results for the growth of axion perturbations, exploring the dependence on the mass of the axion, the temperature scale $\Tl$, and the reheating temperature. In Section~\ref{sec:mini} we discuss the formation of axion miniclusters. We also apply our framework to the QCD axion. We conclude in Section~\ref{sec:conclusions}.

\section{Early matter domination}
\label{sec:EMD}
Before the onset of BBN, the energy density of the Universe is not expected to always be dominated by SM radiation. In fact, it may be governed by a different component, often modeled as a coherently oscillating scalar field, whose energy density redshifts as non-relativistic matter. This leads to an EMD era that departs from the conventional radiation-dominated thermal history~\cite{Allahverdi:2020bys, Batell:2024dsi}. Such a phase can have a significant impact on early-Universe processes, including DM production and the evolution of cosmological perturbations, thereby modifying standard cosmological assumptions~\cite{Steinhardt:1983ia, Lazarides:1990xp, Kawasaki:1995vt, Giudice:2000ex, Grin:2007yg, Visinelli:2009kt, Nelson:2018via, Visinelli:2018wza, Ramberg:2019dgi, Blinov:2019jqc, Arias:2020qty, Carenza:2021ebx, Venegas:2021wwm, Bernal:2021yyb, Arias:2021rer, Bernal:2021bbv}.

In this work, we focus on the time during which the non-relativistic scalar field $\phi$ decays into SM radiation while dominating the energy density of the Universe until the end of the EMD era. The background dynamics is described by the Boltzmann equations~\cite{Giudice:2000ex}
\begin{align}
    \frac{d\rp}{dt} + 3\, H\, \rp &= -\, \Gphi\, \rp\,, \label{eq:boltzmann1}\\
    \frac{d\rR}{dt} + 4\, H\, \rR &= +\, \Gphi\, \rp\,, \label{eq:boltzmann2}
\end{align}
where $\rp$ and $\rR$ denote the energy densities of $\phi$ and SM radiation, respectively, and $\Gphi$ is the decay rate of $\phi$. The Hubble expansion rate $H$ is given by
\begin{equation}
    H^2 = \frac{\rp + \rR}{3\, M_{\rm Pl}^2}\,,
\end{equation}
where $M_{\rm Pl} \simeq 2.4 \times 10^{18}$~GeV is the reduced Planck mass, and the radiation energy density is related to the SM temperature $T$ through
\begin{equation}
    \rR(T) = \frac{\pi^2}{30}\, \gs(T)\, T^4\,,
    \label{rho_r_def}
\end{equation}
with $\gs(T)$ corresponding to the relativistic degrees of freedom that contribute to $\rR$. It is convenient to rewrite Eqs.~\eqref{eq:boltzmann1} and~\eqref{eq:boltzmann2} as a function of the cosmic scale factor $R$ and the comoving variables $\rp\, R^3$ and $\rR\, R^4$ as
\begin{align}
    \frac{d(\rp\, R^3)}{dR} &= -\, \frac{\Gphi}{H\, R}\, (\rp\, R^3)\,, \label{eq:boltzmann1_rew}\\
    \frac{d(\rR\, R^4)}{dR} &= +\, \frac{\Gphi}{H}\, (\rp\, R^3)\,. \label{eq:boltzmann2_rew}
\end{align}

\begin{figure}[t!]
    \centering
    \includegraphics[width=\textwidth]{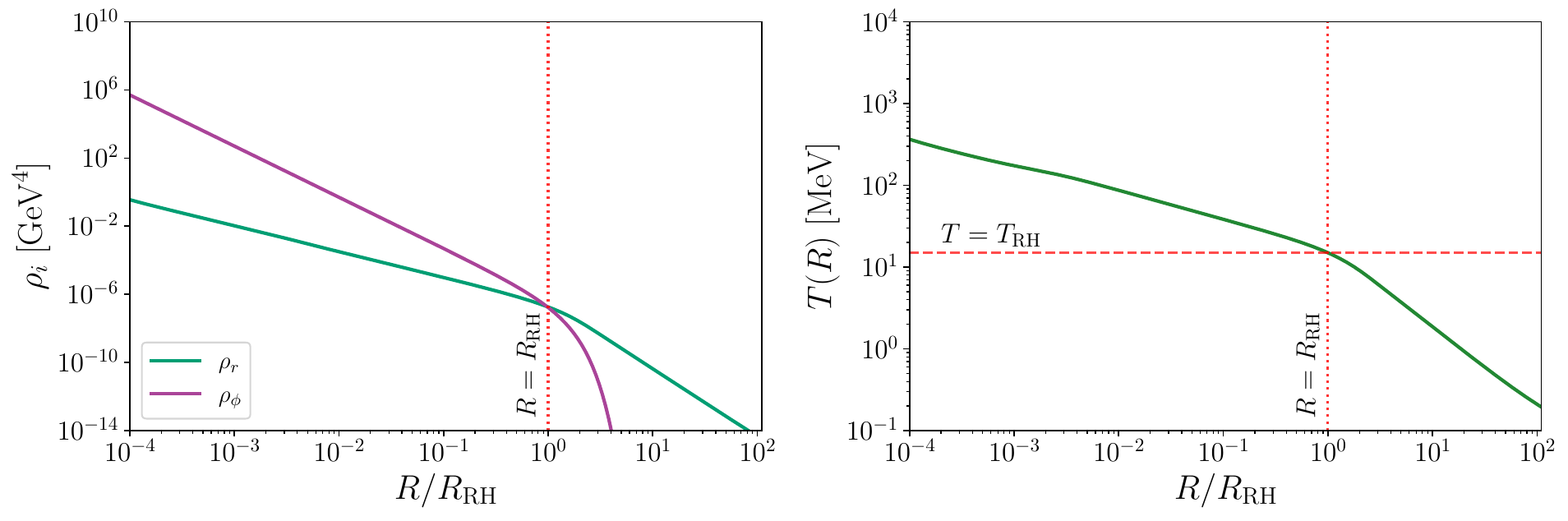}
    \caption{Evolution of the energy densities of the $\phi$ and the SM radiation (left) and the corresponding SM temperature (right), as a function of the cosmic scale factor $R$, for $\TRH = 15$~MeV.}
    \label{fig:background}
\end{figure}

For $R \leq \RRH$, the energy density of the Universe is dominated by $\phi$, inducing EMD cosmology. The scale factor $R = \RRH$ corresponds to the end of the EMD and then to the onset of the standard radiation-dominated phase. The reheating temperature $\TRH \equiv T(\RRH)$ is implicitly defined by the equality $\rp(\TRH) = \rR(\TRH)$. Although we determine it numerically, $\TRH$ can be analytically estimated up to order-one factors through $H_{\rm RH} \equiv H(\TRH) \simeq \Gphi$, which leads to
\begin{equation}
    \TRH^2 \simeq \frac{3}{\pi}\, \sqrt{\frac{10}{\gs(\TRH)}}\,M_{\rm Pl}\, \Gphi\,.
\end{equation}
Consistency with BBN requires the reheating temperature to satisfy $\TRH \gtrsim \Tbbn \simeq 4$~MeV~\cite{Kawasaki:1999na, Kawasaki:2000en, deSalas:2015glj, Hasegawa:2019jsa}. An estimate of the solution of the set of Eqs.~\eqref{eq:boltzmann1_rew} and~\eqref{eq:boltzmann2_rew} can be made analytically, in the regimes where $\phi$ dominates or is subdominant. In these two cases, the Hubble rate evolves as
\begin{equation}
    H(R) \simeq
    \begin{dcases}
        H_{\rm RH}\left(\frac{\RRH}{R}\right)^{3/2} &\text{ for } \Rini \leq R \leq \RRH\,,\\
        H_{\rm RH}\left(\frac{\RRH}{R}\right)^2 &\text{ for } \RRH \leq R\,.
    \end{dcases}
\end{equation}
In addition, one finds that the temperature evolves approximately as~\cite{Giudice:2000ex}
\begin{equation}
    T(R) \simeq
    \begin{dcases}
        \TRH \left(\frac{\RRH}{R}\right)^{3/8} &\text{ for } \Rini \leq R \leq \RRH\,,\\
        \TRH \left(\frac{\RRH}{R}\right) &\text{ for } \RRH \leq R\,,
    \end{dcases}
    \label{eq:temp_scalefact}
\end{equation}
reflecting the continuous injection of entropy from $\phi$ decays into the visible sector during the EMD era. In fact, the decay of the scalar field injects entropy into the radiation bath, diluting any preexisting relic abundances. The left panel of Fig.~\ref{fig:background} shows the evolution of the $\phi$ and SM radiation energy densities as a function of the scale factor for $\TRH = 15$~MeV, while the right panel shows the corresponding evolution of the SM temperature. During the EMD era, the energy density of $\phi$ scales as non-relativistic matter (that is, $\rp(R) \propto R^{-3}$), while the energy density of the radiation scales as $\rR(R) \propto R^{-3/2}$ due to the source term. After the end of the EMD era (when $R = \RRH$), $\phi$ decays exponentially fast and SM radiation scales as free radiation $\rR(R) \propto R^{-4}$. In the upcoming analysis, we assume that the maximum temperature reached by the thermal bath is much larger than the relevant scale for axion DM production, so that the phenomenology is independent of the cosmic initial conditions. Therefore, the only relevant parameter is $\Gphi$ or, equivalently, $\TRH$.

\section{The homogeneous axion} \label{sec:background}
We will focus on the production of axion DM via the standard misalignment mechanism, assuming a pre-inflationary PQ breaking scenario.\footnote{The corresponding isocurvature constraint on the inflationary scale is discussed in Appendix~\ref{sec:iso}.}

Axions are initially frozen at their misalignment value due to Hubble friction in the expanding Universe. As the Hubble rate drops below the axion mass, the field begins coherent oscillations around the minimum of its potential, acting as a cold DM condensate. We focus on the scenario in which axions start oscillating during the EMD era, while their mass is still temperature dependent. In this case, we consider the regime $\Rosc < \RRH$ and $\Rosc < R_\Lambda$, where $\Rosc$ and $R_\Lambda$ denote the scale factors at the onset of oscillations and at the time when the axion reaches its zero-temperature mass, respectively. We now discuss the main features of the background solution in this regime.

After spontaneous breaking of the global symmetry, the axion sector is  described as a minimally coupled real scalar field $a = a(t,\vec{x})$ with action
\begin{equation}
    S_a= \int d^4x\, \sqrt{-g} \left[-\frac12\, (\partial a)^2 - V_a(a)\right],
    \label{eq:ax_action}
\end{equation}
with
\begin{equation}
    V_a(a) \equiv \ma^2(t)\, f_a^2 \left[1 - \cos\left(\frac{a}{f_a}\right)\right],
\end{equation}
where $f_a$ is the scale at which the PQ symmetry is spontaneously broken and $\ma(t)$ is the time-dependent axion mass.

For the QCD axion, $f_a$ determines the axion mass through the topological susceptibility of QCD, $\chi(T)$, as 
\begin{equation}
    \ma^2(T) = \frac{\chi(T)}{f_a^2}\,,
\end{equation}
where from  lattice QCD simulations has been found a zero-temperature value of $\chi(0) \simeq 0.0245$~fm$^{-4}$, in the symmetric isospin case~\cite{Borsanyi:2016ksw}. Therefore, the zero-temperature mass $\maz \equiv \ma(T \to 0)$ of the QCD axion is given by
\be \label{eq:ma}
    \maz \simeq 5.69~\mbox{meV} \left(\frac{10^{9}\,\mbox{GeV}}{f_a} \right).
    \ee

\begin{figure}[t!]
    \centering
    \includegraphics[width=\textwidth]{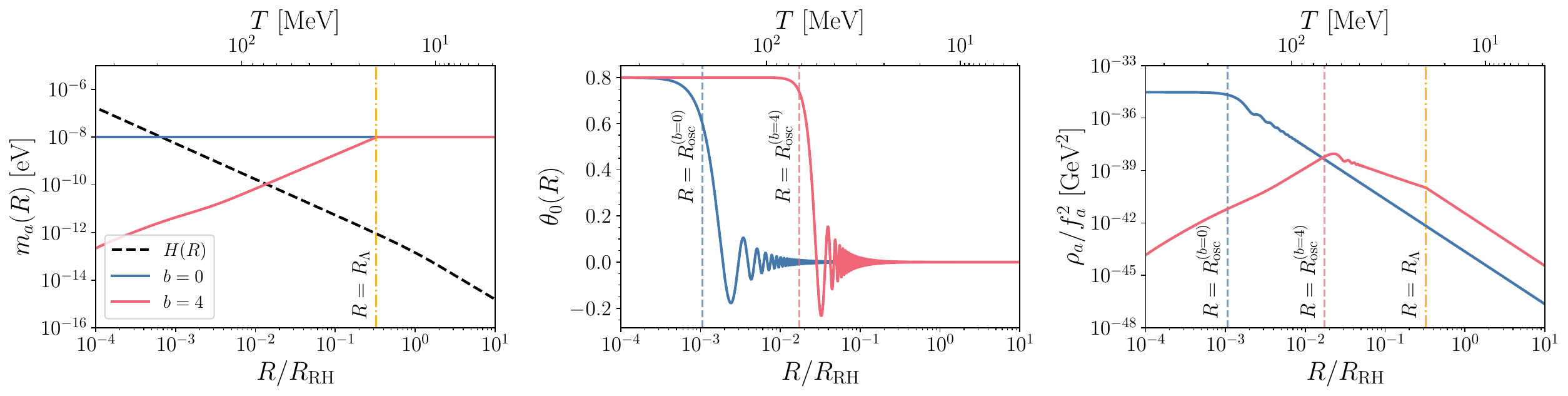}
    \caption{Evolution of the axion mass $\ma$ and the Hubble parameter $H$ (left), the normalized homogeneous axion field $\theta_0$ (center), and the energy density of the axion (right), as functions of the scale factor $R$, for $b = 0$ (blue),  and $b = 4$ (red). We assumed $\TRH = 15$~MeV, $\maz = 10^{-8}$~eV, $\Tl = 25$~MeV, and $\thini = 1$.}
    \label{fig:axion_dynamics}
\end{figure}

Generally speaking, ALPs\footnote{We will loosely call axions to QCD axions and ALPs, and only make the distinction when necessary.} could also feature a temperature dependent mass if the hidden sector breaks non-perturbatively the extra $U(1)$ global symmetry associated with the axion origin. Therefore, for analytical estimations it is useful to consider an approximate expression, which has to be cut off by hand once the mass reaches its zero-temperature value
\begin{equation}
    \ma (T) = \maz \times
    \begin{dcases}
        (\Tl / T)^b &\text{ for } T \geq \Tl, \\
        1 &\text{ for } T \leq \Tl.
    \end{dcases}
    \label{eq:axion_mass}
\end{equation}
For hidden-sector ALPs, $T$ should be understood as the temperature of the sector responsible for generating the axion mass; in the phenomenological analysis below, we identify this temperature with the SM radiation temperature or assume that the two are proportional. From now on, when considering generic ALPs, we will use the prescription of Eq.~\eqref{eq:axion_mass}, considering $b=4$ as a benchmark. Specifically referring to the QCD axion, we will adopt the mass parameterization of Eq.~\eqref{eq:ma}, which follows the lattice results from Ref.~\cite{Borsanyi:2016ksw}. During the EMD era, $\ma(R)\propto R^{3b/8}$, as shown in the left panel of Fig.~\ref{fig:axion_dynamics}.

The equation of motion that follows from Eq.~\eqref{eq:ax_action} is 
\begin{equation}
    \ddot{a} + 3\, H \,\dot a -\frac{\nabla ^2 a}{R^2}+ m_a^2\, f_a\, \sin \left(\frac{a}{f_a}\right) = 0\,,
    \label{eq:eom_axion}
\end{equation}
where the dots ($\dot{}$) denote derivatives with respect to time $t$. 

In the pre-inflationary scenario, the axion field is homogeneous with tiny inhomogeneous perturbations
\begin{equation}
    \theta(t,\vec{x})  \equiv \frac{a(t,\vec x)}{f_a} = \theta_0(t)+\delta\theta(t,\vec{x}) \qquad \text{with} \qquad \delta\theta(t,\vec{x}) \equiv \int \frac{d^3k}{(2\pi)^3} \delta\theta_k(t)e^{i\vec{k}\cdot\vec{x}}\,.
\end{equation}
The homogeneous field $\theta_0$ obeys
\begin{equation}
    \ddot{\theta}_0 + 3\, H\, \dot\theta_0  + \ma^2(t)\, \sin(\theta_0) = 0\,.
    \label{eq:eom_axion_k}
\end{equation}

For a temperature-dependent mass, an approximate WKB analytical solution can be obtained under the adiabatic condition, $\left|\frac{\dot \ma}{\ma^2}\right| \ll 1$, and assuming that the field is already oscillating, i.e., $\ma(t) \gg H(t)$.
In this regime, the solution can be written as
\begin{equation}
    \theta_0(R) \simeq \thini \left(\frac{\Rosc}{R}\right)^{3/2} \sqrt{\frac{\ma(\Tosc)}{\ma(T)}}\, \cos\left(\psi(t) \right),
    \label{WKB_x0}
\end{equation}
where 
$\Tosc \equiv T(\Rosc)$, $\thini \equiv \theta_0(T \gg \Tosc)$ is the initial misalignment angle of the field, and the phase is defined by
\begin{equation}
    \psi(t) \equiv \int^t_{t_\text{osc}} dt'\, \ma(t')\,.
\end{equation}

The temperature $\Tosc$ at which the axions start to oscillate can be estimated by the equality $2\, H(\Tosc) = \ma(\Tosc)$, and corresponds to
\begin{equation}
    \Tosc^{4+b} \simeq \left(\frac{45}{2\pi^2\, \gs(\TRH)}\right)^{1/2} \maz\, M_{\rm Pl}\, \Tl^b\, \TRH^2\,.
\end{equation}
Interestingly, the conditions for oscillations to start during the EMD era, $\TRH < \Tosc$, and during the period where the mass is time dependent, $\Tl < \Tosc$, translate, respectively, into 
\begin{align}
    \frac{\maz}{H_{\rm RH}} &\gtrsim \left(\frac{\TRH}{\Tl}\right)^b,     \label{eq:condition_Tosc} \\
    \frac{\maz}{H_{\rm RH}} &\gtrsim \left(\frac{\Tl}{\TRH}\right)^4.
    \label{eq:condition_Tlambda}
\end{align}
In the following analysis, both conditions are always imposed; however, the hierarchy between $\TRH$ and $\Tl$ is not yet fixed. In addition, the scale factor $\Rosc \equiv R(\Tosc)$ normalized to the scale factor $R_0$ at present is
\begin{equation}
    \frac{\Rosc}{R_0} = \frac{\Rosc}{\RRH}\frac{\RRH}{R_0} \simeq \left(\frac{\TRH}{\Tosc}\right)^{8/3} \left(\frac{\gss(T_0)}{\gss(\TRH)}\right)^{1/3} \frac{T_0}{\TRH}\,.
\end{equation}

The numerical solution of Eq.~\eqref{eq:eom_axion_k} is shown in the central panel of Fig.~\ref{fig:axion_dynamics}. For higher values of $b$, the oscillations are delayed, have a higher frequency, and a faster decay of the envelope. This behavior can be understood from the WKB solution in Eq.~\eqref{WKB_x0}, which yields $\psi(R) \propto R^{(3b+12)/8}$ and $\theta_0(R) \propto R^{-3(8+b)/16}$. In particular, for $b = 4$ the amplitude decays as $R^{-9/4}$, compared to the standard scaling $R^{-3/2}$ in the constant-mass case. As a consequence, the onset of oscillations is delayed for $b > 0$ relative to the constant-mass case ($b = 0$), as shown in the left panel of Fig.~\ref{fig:axion_dynamics}.

In the limit $a/f_a\ll1$, the energy density $\rho_a$ and the pressure $p_a$ of the zeroth mode are given by
\begin{align}
    \rho_a &\simeq \frac12\, f_a^2 \left( \dot{\theta}_0^2 + \ma^2(T) \, \theta_0^2\right), \label{eq:rho_a} \\
    p_a &\simeq \frac12\, f_a^2 \left( \dot{\theta}_0^2 -  \ma^2(T)\, \theta_0^2 \right). \label{eq:p_a}
\end{align}
The time average of the energy density gives $\langle\rho_a\rangle \simeq \frac12\, f_a^2\, \ma^2(T)\, A_\theta^2$, where $A_\theta$ denotes the oscillation envelope of $\theta_0$, which leads to
\begin{equation}
    \langle\rho_a(R)\rangle \propto \left(\frac{\Rosc}{R}\right)^{\frac{24-3b}{8}}.
    \label{eq:rho_a_EMD}
\end{equation}
The evolution of the axion energy density for $b = 0$ (blue line) and $b = 4$ (red line) is shown in the right panel of Fig.~\ref{fig:axion_dynamics}. Once the mass reaches a constant value at $R = R_\Lambda$ (yellow dot-dashed line), the energy density starts to redshift as non-relativistic matter, $\rho_a(R) \propto R^{-3}$.

From the adiabatic conservation of the axion number, the axion energy density at present, $T = T_0$, is
\begin{align}
    \rho_a(T_0) &= \rho_a(\Rosc)\, \frac{\maz}{\ma(\Tosc)} \left(\frac{\Rosc}{R_0}\right)^3 \nonumber\\
    &= \frac{\thini^2\, f_a^2\, T_0^3}{2\, M_{\rm Pl}}\, \frac{\gss(T_0)}{\gss(\TRH)} \left(\frac{2\pi^2\, \gs(\TRH)}{45}\right)^{\frac{8+b}{2(4+b)}} \left(\frac{\maz^b\, \TRH^{4+3b}}{M_{\rm Pl}^4\, \Tl^{4b}}\right)^\frac{1}{4+b}.
    \label{eq:abundance_today}
\end{align}
In particular, for $b = 4$ the axion relic abundance is given by
\begin{equation}
    \Omega_a h^2 \equiv \frac{\rho_a(T_0)}{\rho_c}\, h^2 \simeq 0.12 \left(\frac{\maz}{10^{-6}~\mbox{eV}}\right)^\frac12 \left(\frac{\TRH}{\Tl}\right)^2 \left(\frac{f_a\, \thini}{1.2\times10^{13}~\mbox{GeV}}\right)^2 \left(\frac{10.75}{\gss(\TRH)}\right)^\frac14,
    \label{eq:relic_b4}
\end{equation}
where $\rho_c/h^2 \simeq 1.054 \times 10^{-5}$~GeV/cm$^3$, and the total DM relic abundance measured by Planck is $\Omega_{\rm DM} h^2 \simeq 0.12$~\cite{Planck:2018vyg}. The scale $f_a$ could be tuned to fit the entire observed axion DM abundance in the case of ALPs, while for QCD axions, relation~\eqref{eq:ma} should be used. Smaller values for the energy density are also viable in the case of multi-component DM scenarios.

\section{Perturbation growth during early matter domination}
\label{sec:perturbations}
Having established the background expansion of the Universe and the evolution of the homogeneous axion field, we now turn to the study of the perturbations of the different components during the EMD era.

\subsection{Radiation and metric perturbations}
To study the evolution of perturbations, we need to construct a set of perturbed equations for each of the relevant components; a careful  derivation of the equations can be found in the Appendix~\ref{A_perts}. We start defining the perturbed metric in the Newtonian gauge,
\begin{equation}
    ds^2 = - (1 + 2\, \Phi)\, dt^2 + R^2(t)\, (1 - 2\, \Psi)\, \delta_{ij}\, dx^i\, dx^j,
    \label{eq:metric_newtonian}
\end{equation}
where $\Phi$ and $\Psi$ are the Newtonian potential and the spatial curvature perturbation, respectively~\cite{Dodelson:2003ft}. In the following, we neglect the anisotropic stress, so we set $\Psi = \Phi$. Secondly, we introduce the energy overdensity and velocity divergence for each fluid component $\sigma$ (with $\sigma = r$ for the SM radiation or $\sigma = \phi$), as
\begin{equation}
    \delta_\sigma \equiv \frac{\delta\rho_\sigma}{\rho_\sigma}, \qquad 
    \theta_\sigma \equiv \frac{1}{R}\nabla^2 v_\sigma,
    \label{eq:fluid_perturbed_main}
\end{equation}
where $\delta\rho_\sigma$ denotes the first-order perturbation of the energy density and $v_\sigma$ is the corresponding velocity potential. Density and velocity perturbations are governed by the Einstein equations and the covariant conservation of the stress--energy tensor in a perturbed FLRW space-time. There is energy exchange between $\phi$ and the visible sector, and therefore
\begin{equation} \label{eq:Conservation}
    \nabla_\mu\, {T^{(\sigma)}}^\mu_\nu = Q^{(\sigma)}_\nu \,,
\end{equation}
where $T^{(\sigma)}_{\mu\nu}$ is the stress--energy tensor of the $\sigma$ fluid and $Q^{(\sigma)}_\nu$ encodes the energy--momentum transfer between components. As a result, the stress--energy tensor of each component is not conserved. However, the sum over all species satisfies the covariant conservation law of the total energy--momentum tensor
\begin{equation}
    \sum_{\sigma=\phi,r} \nabla_\mu\, {T^{(\sigma)}}^\mu_\nu = 0 \,.
\end{equation}
The interaction terms are given by~\cite{Erickcek:2011us}
\begin{align}
    Q^{(\phi)}_\nu &= \Gphi\, T^{(\phi)}_{\mu\nu}\, u^\mu\,, \label{Q_terms}\\
    Q^{(r)}_\nu &= -Q^{(\phi)}_\nu \,.
\end{align}
These terms can be expanded at linear order, and their first-order perturbations are given by
\begin{align} 
    \delta Q_0^{(\phi)} &=+\Gphi \, \rp \left( \delta_\phi + \Phi \right), \label{Q_0_c_2} \\
    \delta Q_i^{(\phi)} &=-\Gphi \,\rp \,\partial_i v_{\phi}\,. \label{Q_i_c_2}
\end{align}

With all of the above, we obtain the set of perturbed equations for $\phi$ and the SM radiation:
\begin{align}
    \delta'_\phi &= -\frac{\Gphi}{R\, H} \Phi - \frac{1}{R^2\, H} \theta_\phi + 3\,\Phi' ,\label{eq:deltaphi}\\
    \delta'_r &= \frac{\rp}{\rR}\, \frac{\Gphi}{R\, H}\, (\delta_\phi - \delta_r + \Phi ) - \frac43\, \frac{1}{R^2\, H}\, \theta_r + 4\,\Phi', \label{eq:deltar}\\
    \theta'_\phi &= \frac{k^2}{R^2\, H}\, \Phi - \frac{1}{R}\, \theta_\phi ,\label{eq:thetaphi}\\
    \theta'_r &= \frac{\rp}{\rR}\, \frac{\Gphi}{R\, H} \left( \frac34\, \theta_\phi - \theta_r \right) + \frac{k^2}{R^2\, H} \left( \frac{\delta_r}{4} + \Phi \right), \label{eq:thetar}\\
    \Phi' &= -\left[ \frac{1}{6\, M_\text{Pl}^2\, R\, H^2}\, (\rp\, \delta_\phi + \rR\, \delta_r + \rho_a\, \delta_a) + \left( \frac{k^2}{3\, R^3\, H^2} + \frac{1}{R} \right) \Phi \right] .\label{eq:phi}
\end{align}
We note that the axion appears in the Hubble parameter and in the Poisson Eq.~\eqref{eq:phi}, but its contribution is small and can be safely neglected. Thus, our strategy is to solve the above set of equations ignoring the axion contribution and later use it as input to solve for its evolution. The initial conditions used are $\theta_\phi(\Rini) = \theta_r(\Rini) = 0$, $\delta_\phi(\Rini) = -2\, \Phi_\text{p}$, $\delta_r(\Rini) = -\Phi_\text{p}$, and $\Phi(\Rini) = \Phi_\text{p} \equiv \sqrt{\mathcal{A}_s}$ with $\mathcal{A}_s \simeq 2.101 \times 10^{-9}$ being the measured amplitude of the primordial scalar power spectrum~\cite{Planck:2018vyg}. Even if in the following a full numerical treatment is presented, in Appendix~\ref{sec:analytics} analytical solutions for the evolution of the perturbations are presented.

\begin{figure}[t!]
    \centering
    \includegraphics[width=\textwidth]{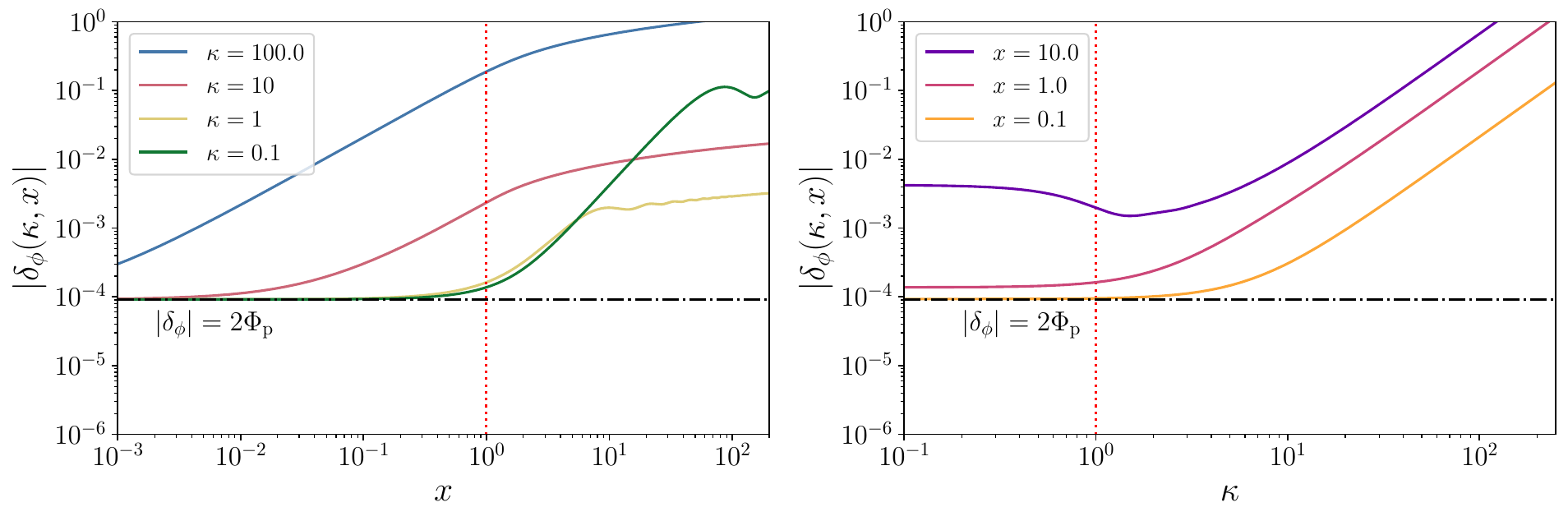}
    \includegraphics[width=\textwidth]{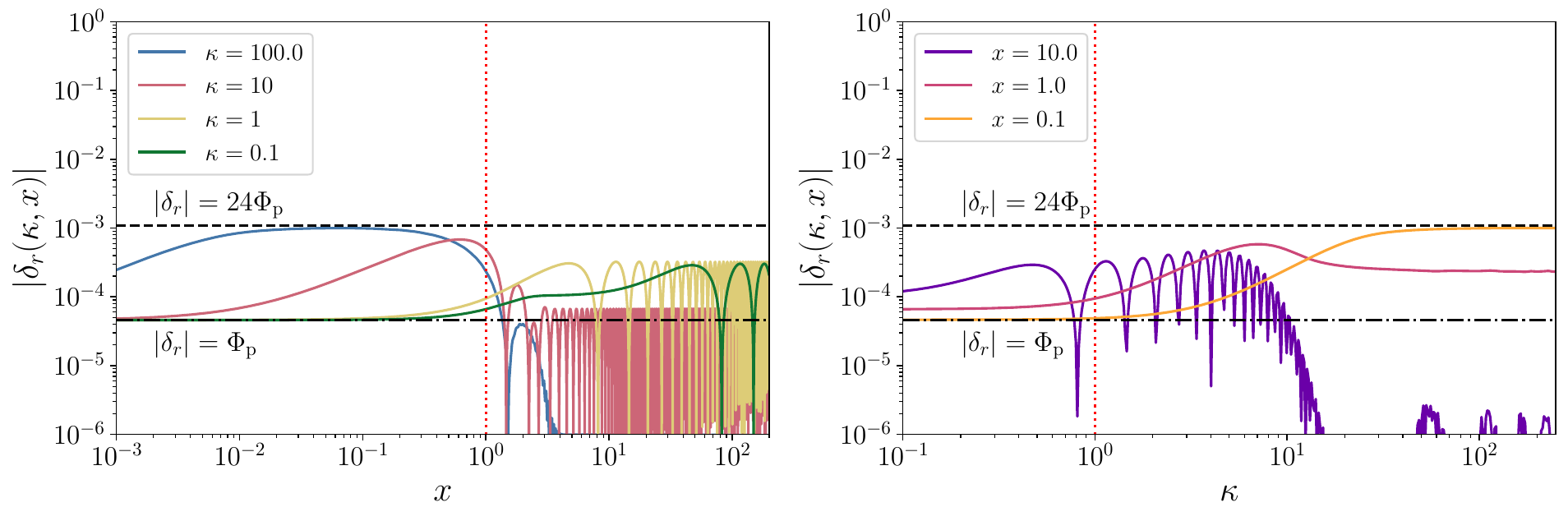}
    \caption{Evolution of the absolute value of the overdensities for $\phi$ (top) and the SM radiation (bottom) as a function of the scale factor (left) and the mode (right).}
    \label{fig:delta}
\end{figure}
Figure~\ref{fig:delta} shows the evolution of the absolute value of the overdensities for $\phi$ (top) and the SM radiation (bottom) as a function of the scale factor (left) and the mode (right), where we have conveniently introduced a set of dimensionless quantities, defined as
\begin{equation}
    x \equiv \frac{R}{\RRH}, \qquad \kappa \equiv \frac{k}{\kRH}, \qquad \text{with} \qquad \kRH \equiv \RRH\, H_{\rm RH}\,.
    \label{eq:new_vars}
\end{equation}
It is interesting to note that the choice of these variables makes the figure independent of $\TRH$. During the EMD era (that is, $x < 1$), modes for the $\phi$ overdensity with $\kappa > 1$ enter the horizon before reheating and grow linearly as the Universe is dominated by non-relativistic matter. It is interesting to note that the nonlinear regime for the $\phi$ overdensity (that is, $|\delta_\phi| \geq 1$) can only be reached for modes corresponding to $\kappa \gtrsim \sqrt{3/(2\, \Phi_\text{p})} \simeq 181$; see Appendix~\ref{sec:analytics}. After the EMD era ($x > 1$) the growth is only logarithmic since the Universe has transitioned into a radiation-dominated era. The situation is different for radiation overdensities as they have a non-vanishing pressure, which implies that their linear growth during the EMD era saturates once it reaches an enhancement of order 24, see Appendix~\ref{sec:analytics}. After the EMD era, $|\delta_r|$ oscillates with a constant amplitude.

\begin{figure}[t!]
    \centering
    \includegraphics[width=\textwidth]{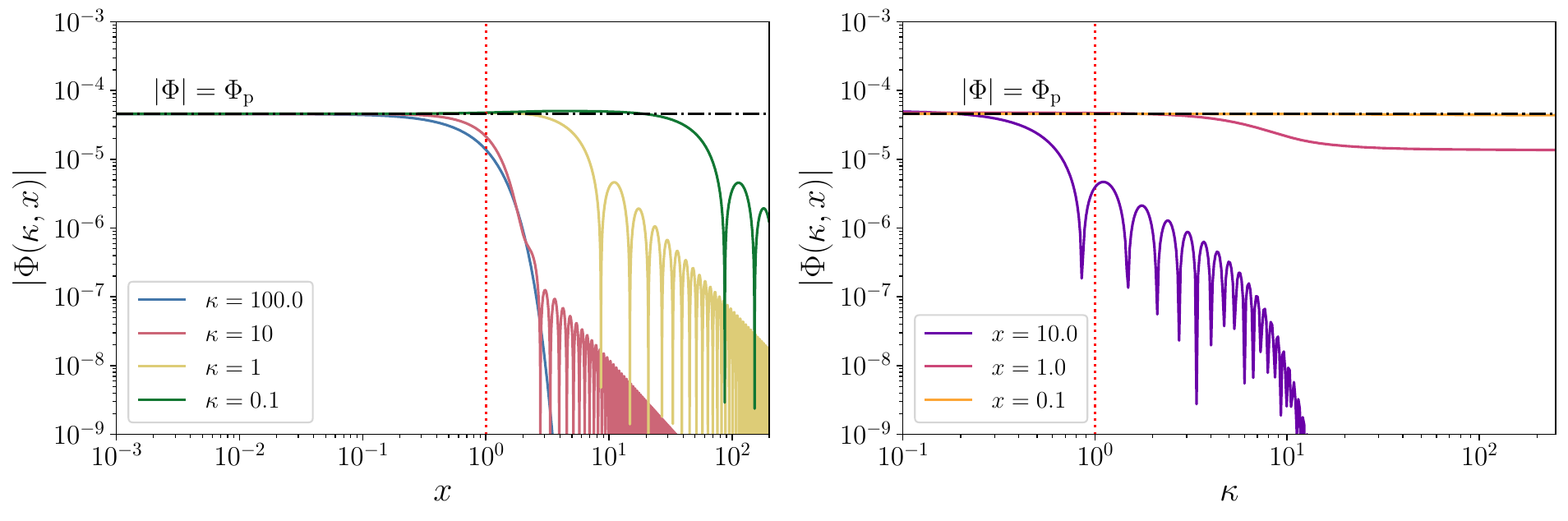}
    \caption{The evolution of the absolute value of the gravitational potential $\Phi$.}
    \label{fig:Phi}
\end{figure}
Finally, the evolution of the absolute value of the gravitational potential $\Phi$ is shown in Fig.~\ref{fig:Phi}. Even if during the EMD era it remains constant, after the EMD era it rapidly decreases through damped oscillations due to the progressive conversion of $\phi$ into SM radiation.

\subsection{Axion perturbations}
We now turn to the evolution of axion perturbations within the framework described above. In principle, one could describe the axion sector as a perturbed fluid. However, we aim to include the temperature-dependent mass in Eq.~\eqref{eq:axion_mass}, which induces an explicit energy exchange between radiation and axions, since the mass itself is perturbed:
\begin{equation}
    \ma^2(T) = \ma^2(\overline T) + \delta \ma^2 \qquad \text{with} \qquad \delta \ma^2 = \frac{d\ma^2}{dT}\,\delta T_k\,,
\end{equation}
where $\delta T_k$ is the linear temperature perturbation with momentum $k$ and $\overline T$ is the background photon temperature. This makes the fluid description cumbersome, as the perturbed mass introduces a direct coupling between the axion and radiation perturbations and breaks the usual conservation structure. Therefore, we follow the axion perturbations directly from its field equations. These will be solved using the solution of Eqs.~\eqref{eq:deltaphi} to~\eqref{eq:phi} as input.

We start by considering a perturbation of the axion field, parameterized as the evolution of the misalignment angle
\begin{equation}
    \theta(t,\vec{x}) = \theta_0(t)+\delta\theta(t,\vec{x}) \qquad \text{with} \qquad \delta\theta(t,\vec{x}) = \int \frac{d^3k}{(2\pi)^3} \delta\theta_k(t)e^{i\vec{k}\cdot\vec{x}}\,,
    \label{expantion_field}
\end{equation}
where $\theta_0(t)$ corresponds to the homogeneous solution discussed in the previous section, and $\delta \theta_k(t)$ is the linear perturbation with momentum $k$. The equation of motion for the perturbation follows from the variation of the action in Eq.~\eqref{eq:ax_action}, and is given by
\begin{equation}
    \frac{d^2 \delta \theta_k}{dt^2} + 3 H\, \frac{d \delta\theta_k}{dt} + \frac{k^2}{R^2} \delta \theta_k + \ma^2\, \cos\theta_0\, \delta \theta_k = 4 \dot \Phi\, \dot \theta_0 - 2 \Phi\, \ma^2\, \sin\theta_0 -\frac14\, \frac{d\ma^2}{dT} \sin\theta_0\, \overline{T}\, \delta_r\,,
    \label{eq:perturbation_t}
\end{equation}
where we have used $\delta_r = 4\, \delta T/\overline{T}$ (ignoring the change in the effective relativistic degrees of freedom); see Appendix~\ref{B_perts}. 
Equation~\eqref{eq:perturbation_t} can be rewritten as
\begin{equation}
    \frac{d^2 \delta\theta_\kappa}{dx^2} + \left[\frac{d\ln(H\, x)}{dx} + \frac{3}{x}\right] \frac{d\delta\theta_\kappa}{dx} + \left[\frac{\kappa^2}{x^4}\left(\frac{H_{\rm RH}}{H}\right)^2 + \left(\frac{\ma(x)}{H\,x}\right)^2 \cos\theta_0(x)\right] \delta\theta_\kappa = S_1 + S_2 + S_3\,,
    \label{eq:eq_axion_pert}
\end{equation}
where the source terms are given by
\begin{align}
    S_1 &\equiv 4\,\frac{d\Phi}{dx}\,\frac{d\theta_0}{dx}, \label{eq:S1} \\
    S_2 &\equiv -2\,\Phi \left(\frac{\ma(x)}{H(x)\,x}\right)^2 \sin\theta_0(x), \label{eq:S2} \\
    S_3 &\equiv -\frac{d\ln \ma^2}{d\ln T}\, \frac{\delta_r}{4}\, \left(\frac{\ma(x)}{H(x)\,x}\right)^2 \sin\theta_0(x) \simeq
    \begin{dcases}
        \frac{b\, \delta_r}{2}\, \left(\frac{\ma(x)}{H(x)\,x}\right)^2 \sin\theta_0(x) &\text{for } x < x_\Lambda,\\
        0 &\text{for } x > x_\Lambda,
    \end{dcases}
    \label{eq:S3}
\end{align}
and $x_\Lambda \equiv R_\Lambda/\RRH$. We note that for a constant axion mass, the source term $S_3$ vanishes. In the following, Eq.~\eqref{eq:eq_axion_pert} will be numerically solved using the initial conditions $\delta\theta_k(\Rini) = \delta\theta'_k(\Rini) = 0$, since we focus on the adiabatic solution for the axion perturbation, neglecting the subdominant isocurvature mode (cf Appendix~\ref{B_perts}).\footnote{We keep the initial misalignment angle away from $\thini\sim\pi$ to avoid parametric growth of the axion fluctuations~\cite{Arvanitaki:2019rax}; the so-called `large misalignment' scenario. Although this amplification could, in principle, occur within our framework, our primary focus remains on isolating the impact of the mass coupling to radiation.}

To solve the perturbed axion equations we proceed as follows: First, we compute the background evolution of $\phi$ and radiation in Eqs.~\eqref{eq:boltzmann1_rew} and~\eqref{eq:boltzmann2_rew}. We then use this solution to determine the background evolution of the axion zeroth mode, Eq.~\eqref{eq:eom_axion_k}. Next, we solve for the evolution of the metric and radiation perturbations using Eqs.~\eqref{eq:deltaphi} to~\eqref{eq:phi}, and finally compute the axion perturbations following Eq.~\eqref{eq:eq_axion_pert}.

\begin{figure}[t]
    \centering
    \includegraphics[width=\linewidth]{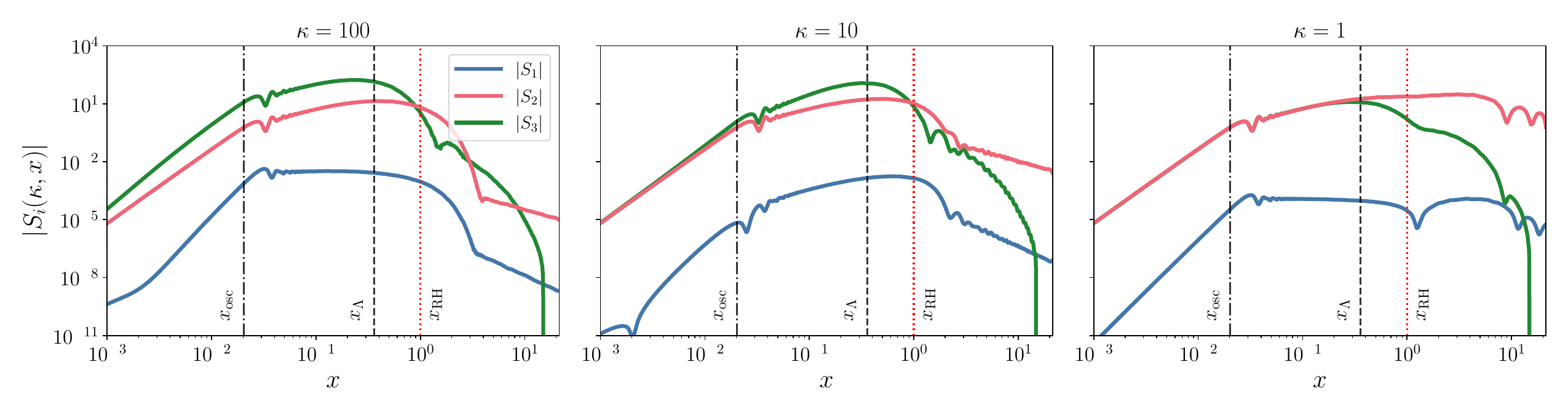}
    \caption{Source terms $S_1$, $S_2$, and $S_3$ as a function of $x$ for different values of $\kappa$, assuming $\TRH = 100$~MeV, $\maz = 2.2\times 10^{-7}$~eV, $\Tl = 150$~MeV, and $\thini = 1$. We consider three representative modes with $\kappa = 100$, $\kappa = 10$, and $\kappa = 1$. We show the envelope of the functions to avoid the noise from oscillations.}
    \label{fig:sources_vs_R_panels}
\end{figure}
The evolution of the three source terms for the QCD axion case is illustrated in Fig.~\ref{fig:sources_vs_R_panels}, for $\TRH = 100$~MeV, $\maz \sim 2.2 \times 10^{-7}$~eV, $\Tl = 150$~MeV and $\thini = 1$. During the EMD era, the gravitational potential is nearly constant and therefore $S_1$ is very suppressed and subdominant with respect to $S_2$ and $S_3$. The last two sources have a similar scaling and their ratio $|S_3/S_2| = b/4 \times |\delta_r/\Phi| \to |\delta_r/\Phi|$. This implies that for superhorizon modes $|S_3/S_2| \simeq 1$, while for subhorizon modes $|S_3/S_2| \simeq 24$. $S_2$ only dominates over $S_3$ in the case where the mass of the axion is constant, since $S_3 = 0$. Additionally, the growth of these sources is driven by the common factor $(\ma/(H\, x))^2 \propto x^{(3b+4)/4}$, which corresponds to $x^4$ or $x^1$ if the axion mass evolves or not, respectively.

\subsection{Axion overdensity evolution}
The axion overdensity is defined as $\delta_a \equiv \delta \rho_a / \rho_a$, where $\delta \rho_a$ denotes the fluctuation of the axion energy density. We first express $\delta \rho_a$ in terms of the perturbed field as (see Appendix~\ref{B_perts})
\begin{equation}
    \delta \rho_a = f_a^2 \left[ \dot{\theta}_0\, \delta\dot{\theta}_k  - \Phi\, \dot{\theta}_0^2 + \ma^2\, \sin{\theta_0}\, \delta\theta_k - \frac{b}{2} \ma^2\, (1 - \cos{\theta_0})\, \delta_r \right].
\end{equation}
Although the axion density is given by Eq.~\eqref{eq:rho_a}, its overdensity is
\begin{equation}
    \delta_a = \frac{\dot{\theta}_0\, \delta\dot{\theta}_k  - \Phi\, \dot{\theta}_0^2 + \ma^2\, \sin{\theta_0}\, \delta\theta_k - \frac{b}{2}\, \delta_r\, \ma^2\, (1 - \cos{\theta_0})}{\frac12 \dot\theta_0^2 + \ma^2\, (1 - \cos\theta_0)}\,,
    \label{eq:overdensity_numeric}
\end{equation}
for $T > \Tl$. For $T < \Tl$, the mass has saturated and the last term in the numerator should be dropped. In the following, the general case will be studied, followed by the specific case of the QCD axion.

\subsubsection{General case}
\begin{figure}[t!]
    \centering
    \includegraphics[width=\textwidth]{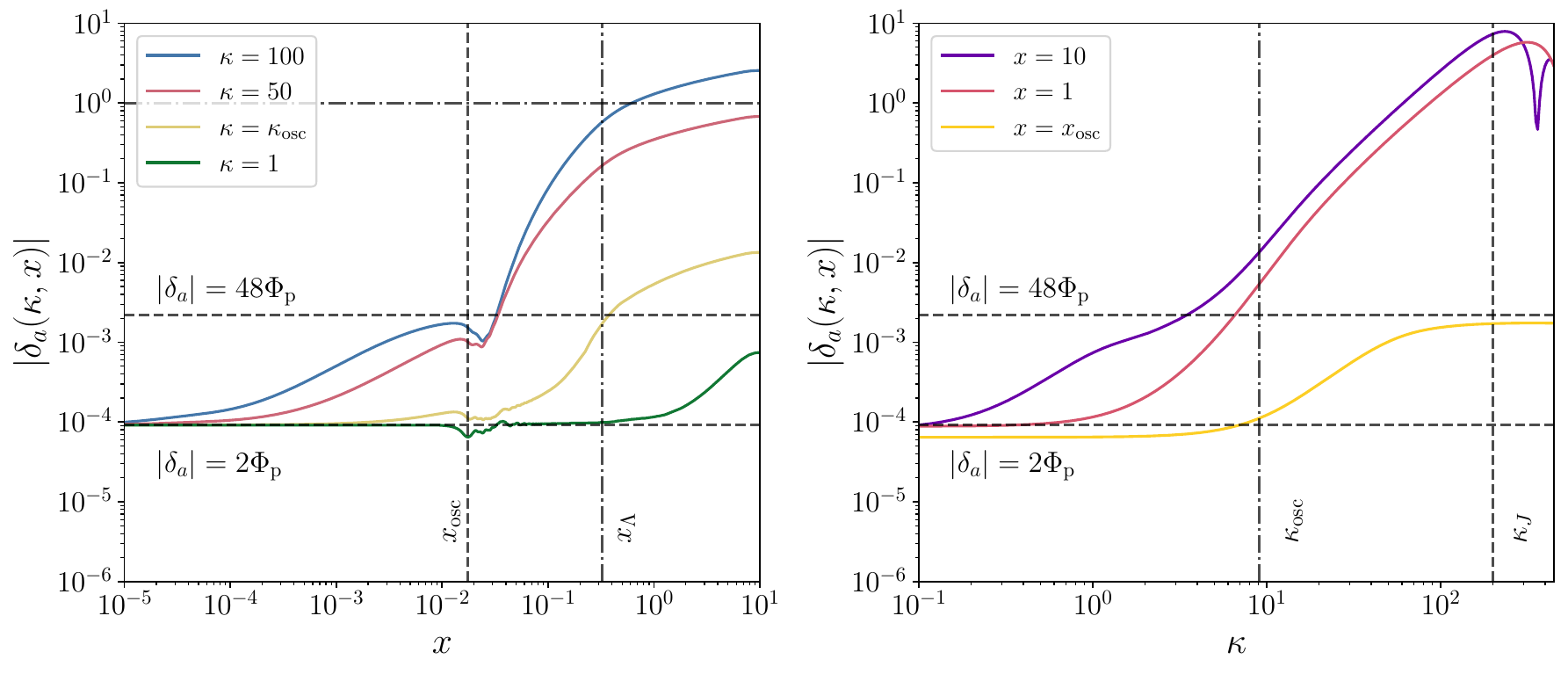}
    \caption{Evolution of the absolute value of the axion overdensity for $\TRH = 15$~MeV, $\maz =10^{-8}$~eV, $\Tl = 25$~MeV, $b = 4$ and $\thini = 1$, as a function of $x$ (left) or $\kappa$ (right).
    }
    \label{fig:deltaa}
\end{figure}
In Fig.~\ref{fig:deltaa} we show the absolute value of the axion overdensity for $\TRH = 15$~MeV, $\maz =10^{-8}$~eV, $\Tl = 25$~MeV, $b = 4$, and $\thini = 1$. The left panel shows $|\delta_a|$ as a function of the normalized scale factor $x$ for several values of $\kappa$, while the right panel shows it as a function of the comoving mode $\kappa$ for different values of $x$. Note that before the onset of axion oscillations $\dot\theta_0 \simeq 0$, and therefore Eq.~\eqref{eq:overdensity_numeric} reduces to
\begin{equation}
    \delta_a \simeq 2\, \frac{\delta\theta_k}{\theta_0} - \frac{b}{2}\, \delta_r \simeq -\frac{b}{2}\, \delta_r\,.
\end{equation}
Consequently, modes that are subhorizon at the onset of axion oscillations reach a maximum amplitude set by the maximum amplitude of $\delta_r$. As seen in Fig.~\ref{fig:delta}, the largest enhancement corresponds to modes that entered the horizon well before oscillations begin ($\kappa\gg1$) and is given by $|\delta_a| = 12\, b\, \Phi_{\rm p}$. In contrast, modes that are superhorizon at reheating ($\kappa \ll 1$) remain frozen, so the overdensity stays close to $|\delta_a| \simeq \frac{b}{2}\, \Phi_\text{p}$. After $x_{\rm osc}$, $|\delta_a|$ inherits a transient oscillation from the background field $\theta_0(t)$ through the trigonometric factors in Eq.~\eqref{eq:overdensity_numeric}, which damp out as the oscillation amplitude decreases and the field enters the small-angle regime, where the potential becomes quadratic and these factors reduce to a common $\theta_0^2(t)$ envelope shared by numerator and denominator. The right panel of Fig.~\ref{fig:deltaa} shows that modes entering the horizon before the onset of oscillations ($\kappa > \kosc$) benefit the most from the energy transfer with radiation and the non-standard cosmological evolution, with the enhancement growing monotonically with $\kappa$. This trend does not extend indefinitely, since for $\kappa \gtrsim \kappa_J$ (dashed line) the quantum pressure dominates and suppresses the growth of the overdensity. The Jeans threshold of modes that entered the horizon before the axion mass became constant is given by
\begin{equation}
    \kappa_J \simeq \sqrt{\frac{\maz}{H_{\rm RH}}}\, x_\Lambda^{1/4} \simeq 91 \left(\frac{\maz}{10^{-8}\,\mbox{eV}}\, \frac{10^{-21}\,\mbox{GeV}}{H_{\rm RH}}\right)^{1/2} \left(\frac{x_{\Lambda}}{0.7}\right)^{1/4},
    \label{eq:Jeans}
\end{equation}
see Appendix~\ref{sec:jeans} for the detailed derivation. The modes with $\kappa \gtrsim \kappa_J$ are therefore suppressed after $x_\Lambda$, in good agreement with our numerical results.

\begin{figure}[t!]
    \centering
    \includegraphics[width=\linewidth]{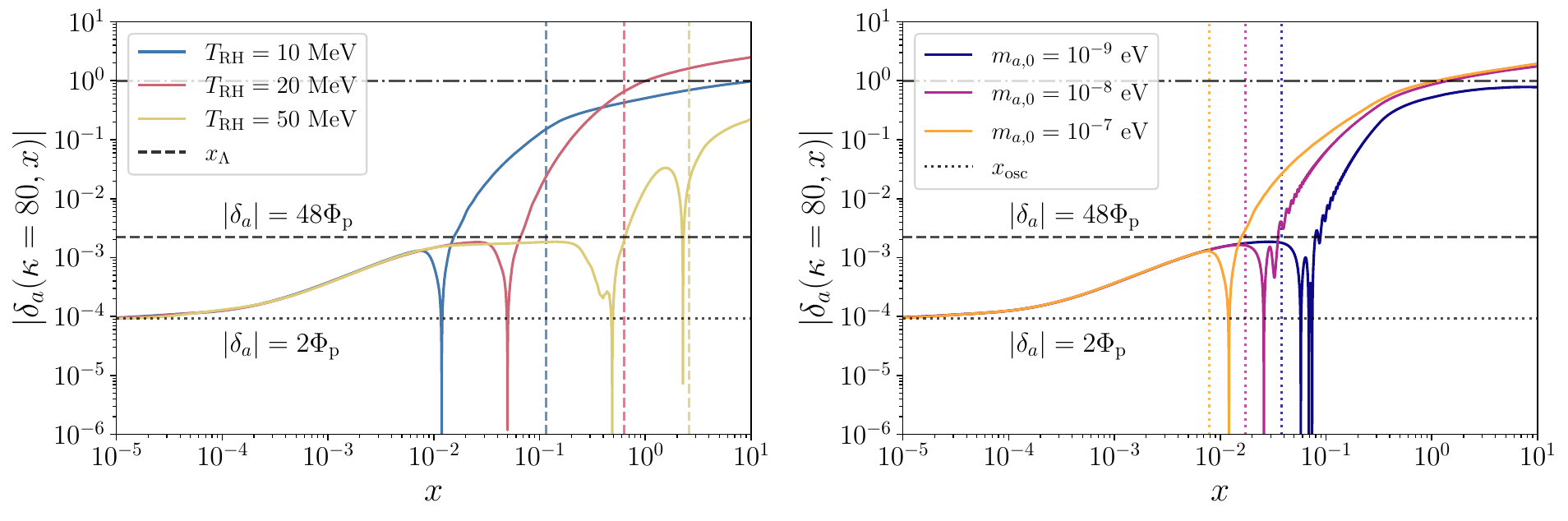}
    \includegraphics[width=\linewidth]{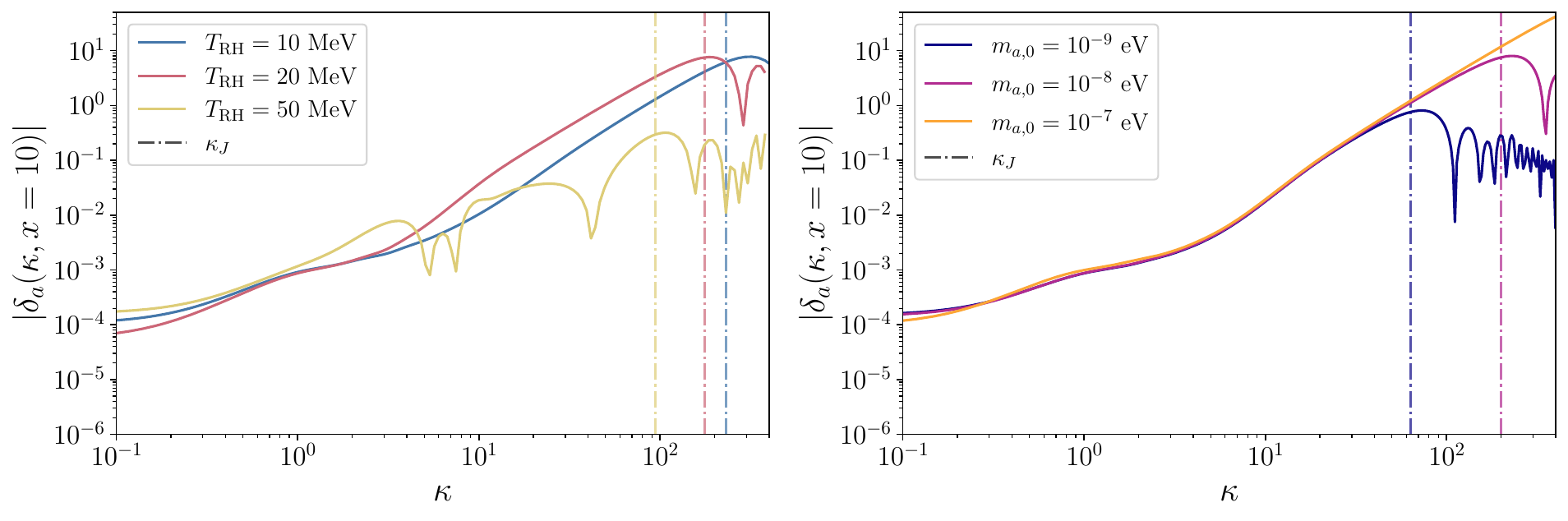}
    \caption{Axion overdensity $|\delta_a|$ for different $\TRH$ (left panels) and $\maz$ (right panels), with $\Tl = 25$~MeV and $b = 4$. The left panels fix $\maz = 10^{-8}$~eV, while the right panels fix $\TRH = 15$~MeV. The top panels show the evolution as a function of $x$ for $\kappa = 80$, and the bottom panels the spectra as a function of $\kappa$ at $x = 10$. The Jeans scale $\kappa_J$ is indicated by vertical dashed lines.}
    \label{fig:deltaa2}
\end{figure}
In Fig.~\ref{fig:deltaa2}, we examine the impact of varying the temperature $\TRH$ (left panels) and the mass $\maz$ (right panels) on the overdensity of the axion, for $\Tl = 25$~MeV, $b = 4$, and $\thini = 1$. In the left panels, where the mass is fixed at $\maz = 10^{-8}$~eV, we find that the overdensity of the axion is maximized for large modes $\kappa$ (but smaller than $\kappa_J$) when $\TRH$ lies slightly below $\Tl$ (e.g. $\TRH = 20$~MeV), ensuring that the temperature dependence of the mass ceases before the end of the EMD era. For $\TRH = 10$~MeV, oscillations begin earlier, so the source $S_3$ acts over a shorter interval and its enhancement is only partially imprinted on the perturbation, yielding a reduced amplitude (top-left panel). The earlier transition to a temperature-independent mass further suppresses the effective duration of $S_3$. When $\Tl < \TRH$ (e.g.\ $\TRH = 50$~MeV), the overdensity tracks the $\delta_r$ plateau throughout EMD but is suppressed near reheating, where $\delta_r$ decays sharply for large modes, yielding only a mild net enhancement. In this regime, the modes retain the large amplitude of the radiation overdensity characteristic of $\kappa \gg 1$, while avoiding the suppression that arises near reheating. Finally, in the right panel of Fig.~\ref{fig:deltaa2}, we fix $\TRH = 15$~MeV and vary $\maz$. Larger masses ($\maz = 10^{-8}$~eV and $\maz = 10^{-7}$~eV) exhibit the strongest enhancements at high $\kappa$, which in fact reach the nonlinear regime around $\kappa \gtrsim 200$. As the axion mass decreases, the amplification is reduced, since higher-$\kappa$ modes become pressure suppressed. The smaller the mass, the lower the threshold wavenumber at which this suppression begins, as shown in Eq.~\eqref{eq:Jeans}. In summary, we find that the optimal scenario for maximizing axion overdensity growth is achieved when $\Tl$ is slightly above but close to $\TRH$, and for large masses of axions.

\begin{figure}[t!]
    \centering
    \includegraphics[width=\textwidth]{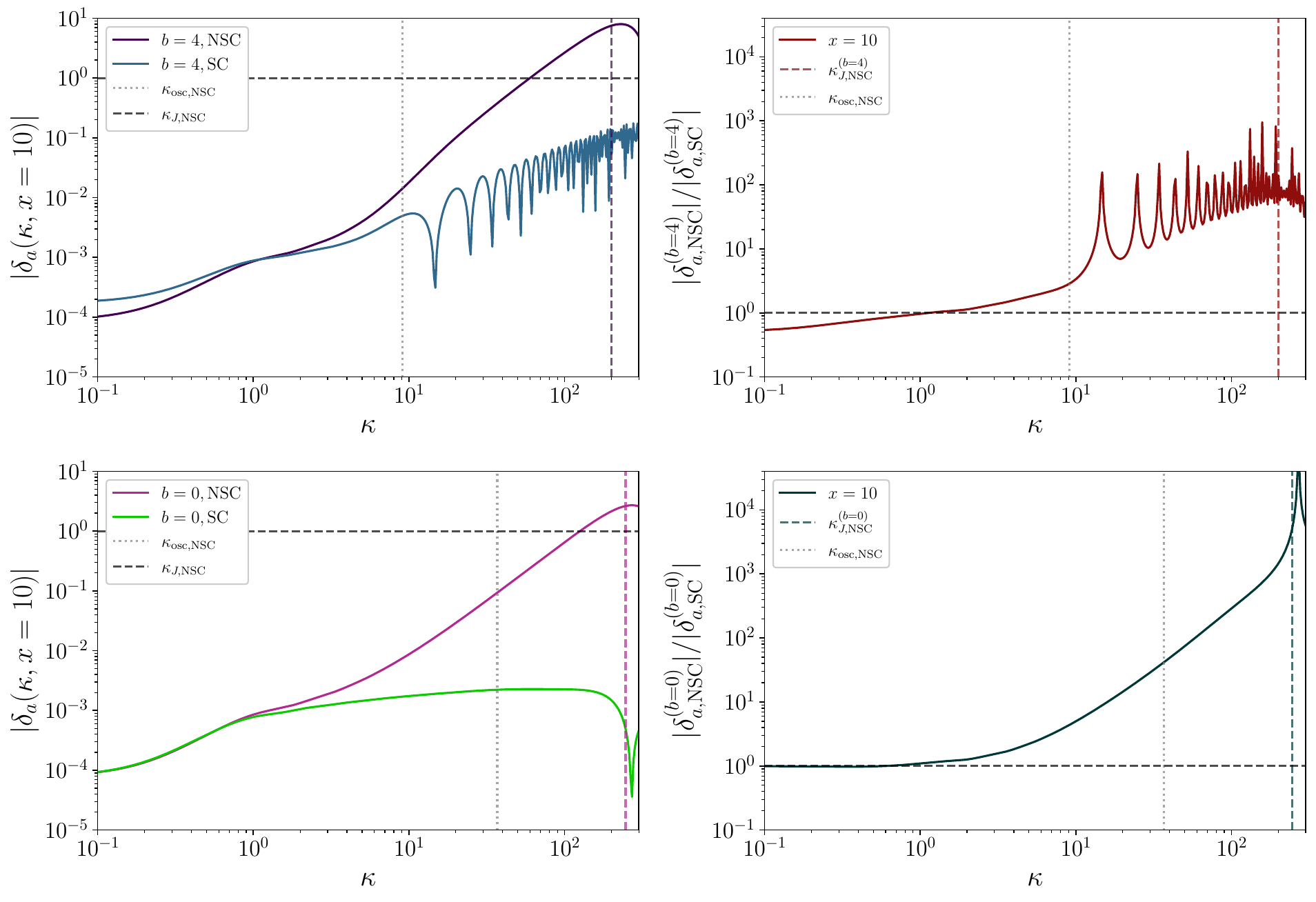}
    \caption{Axion overdensity spectrum in the cases where the reheating temperature is low ($\TRH = 15$~MeV) and high reheating temperature ($\TRH = 8$~TeV) (left panels), and their ratio (right panels). The top (bottom) panels correspond to $b = 4$ ($b = 0$), for $\maz = 10^{-8}$~eV, $\Tl = 25$~MeV, $\thini = 1$, and $x = 10$.}
    \label{fig:comparison}
\end{figure}
In order to understand the effect of the EMD era, in Fig.~\ref{fig:comparison} we show the absolute value of the axion overdensity for $b = 4$ (top) and $b = 0$ (bottom) for the cases where the reheating temperature is high or low (left) and the ratio between them (right), for $\maz = 10^{-8}$~eV, $\Tl = 25$~MeV, $\thini = 1$, and $x = 10$. For the case with low-reheating temperature we assume $\TRH = 15$~MeV, while for the case with high-reheating temperature we take $\TRH = 8$~TeV; however, the dynamics of the axion becomes independent of the reheating temperature as long as $\TRH \gtrsim 10$~GeV. It is interesting to note that subhorizon modes smaller than the Jeans scale at the onset of oscillations benefit from a large enhancement due to the early-matter domination. For $b = 4$ the enhancement is typically of order $10^2$, while for $b = 0$ it could reach $10^3$.

\begin{figure}[t!]
    \centering
    \includegraphics[width=0.9\textwidth]{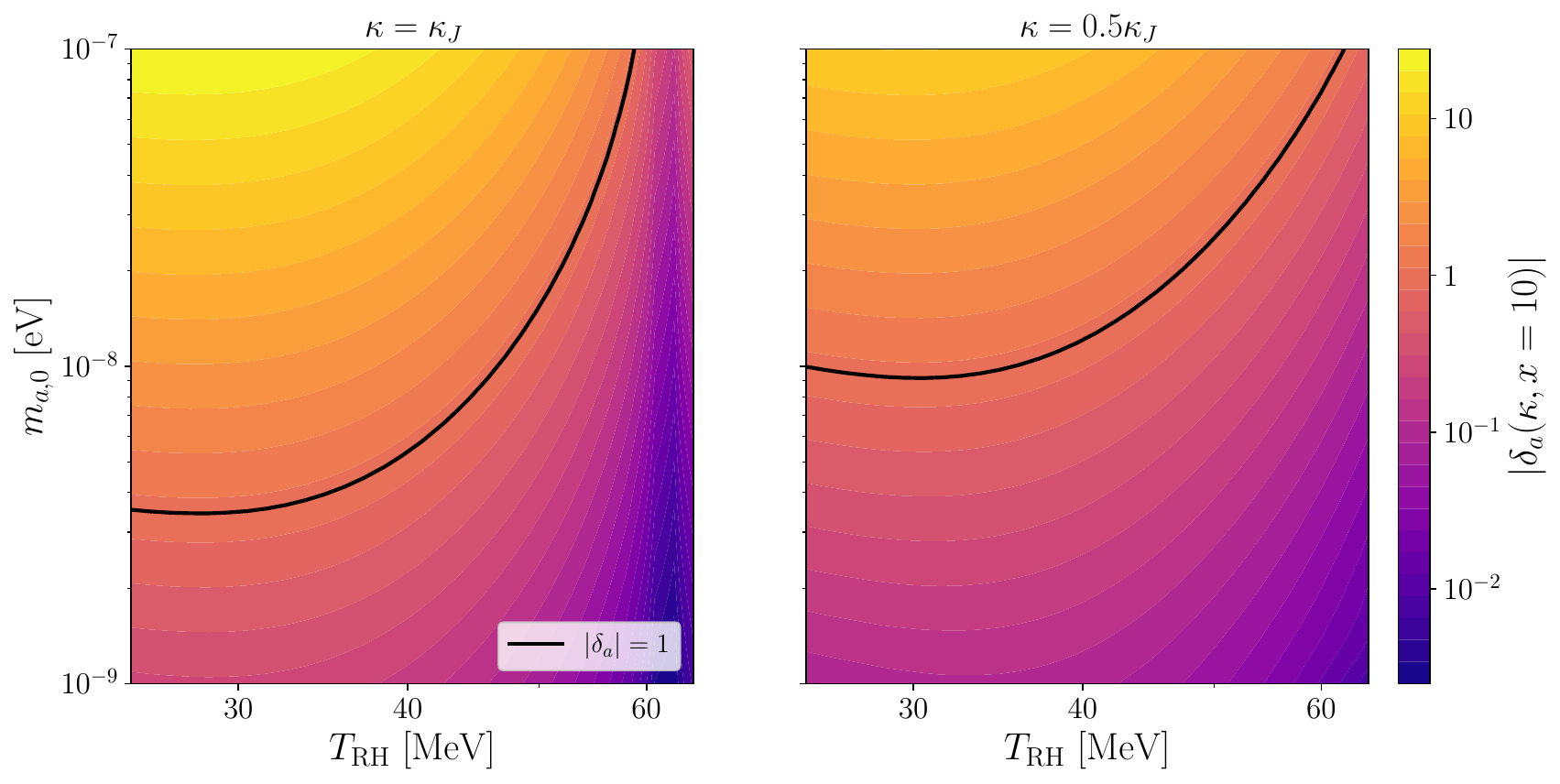}
    \caption{Absolute value of the ALP overdensity $\delta_a$ as a function of the reheating temperature $\TRH$ and the zero-temperature axion mass $\maz$ for $\kappa=\kappa_J$ and $\kappa=0.5\,\kappa_J$, respectively. We have fixed $\thini = 1$, $\Tl=40$~MeV, and $b = 4$. All the parameter space satisfies the correct DM relic abundance. }
    \label{fig:Jeans_grads}
\end{figure}

Finally, in Fig.~\ref{fig:Jeans_grads} we show a gradient map for the ALP overdensity as a function of the reheating temperature and the axion zero-temperature mass, for two different mode scales: $\kappa = \kappa_J$ (left) and $\kappa=0.5~\kappa_J$ (right). As expected, the highest enhancement of the overdensity is obtained for the highest non-suppressed mode, which is the one set by the Jeans threshold of Eq.~\eqref{eq:Jeans}, large values for the axion mass, and $\TRH \lesssim \Tl$. This trend has also been observed in Ref.~\cite{Hardy:2026mkp}, and is due to the source term $S_2$ that bounds the ALP overdensity with the overdensity of the field $\phi$. However, it is interesting to note that modes  $\kappa=0.5\,k_J$ receive comparable enhancements from the $S_3$ source term. The range of mass that can be enhanced is highly dependent on the temperature scale $\Tl$. Smaller masses can be efficiently enhanced if $\Tl$ is slightly above the smallest allowed reheating temperature, i.e., $\Tbbn$.

\subsubsection{The QCD axion case}\label{sec:qcd_axion}
For the QCD axion, the confinement scale is fixed at $\Tl = \TQCD \simeq 150$~MeV. Reaching the nonlinear regime after the EMD era thus requires the optimal hierarchy $\Tl \gtrsim \TRH$ identified above, which selects $\TRH$ in the range $\Tl/\TRH \sim 1.1$--$2$, {\it i.e.} $\TRH$ between 75~MeV and 130~MeV. Imposing $\Omega_a h^2 = \Omega_{\rm DM} h^2$ in Eq.~\eqref{eq:relic_b4} together with the QCD relation in Eq.~\eqref{eq:ma} fixes the axion mass for a given $\thini$ to be
\begin{equation}
    \label{relicdensityconstraint}
    \maz \simeq 3.58 \times 10^{-7}~\mbox{eV} \left(\frac{g_{\star}(\TRH)^{\frac{1}{2}}}{g_{\star, s}(\TRH)^{\frac{2}{3}}}\right)\left(\frac{\TRH}{150~\text{MeV}}\right)^{4/3}\, \thini^{4/3}.
\end{equation}
Therefore, the masses that receive the greatest enhancement of their overdensity perturbations are in the range $1.4 \times 10^{-7}~\mbox{eV} \lesssim \maz \lesssim 3.2 \times 10^{-7}$~eV, for $\thini =1$. For these masses, the overdensity enters the nonlinear regime at or near the EMD era. However, higher and lower axion masses can still reach nonlinearity at or before matter-radiation equality (MRE). 

\begin{figure}[t!]
    \centering
    \includegraphics[width=0.7\linewidth]{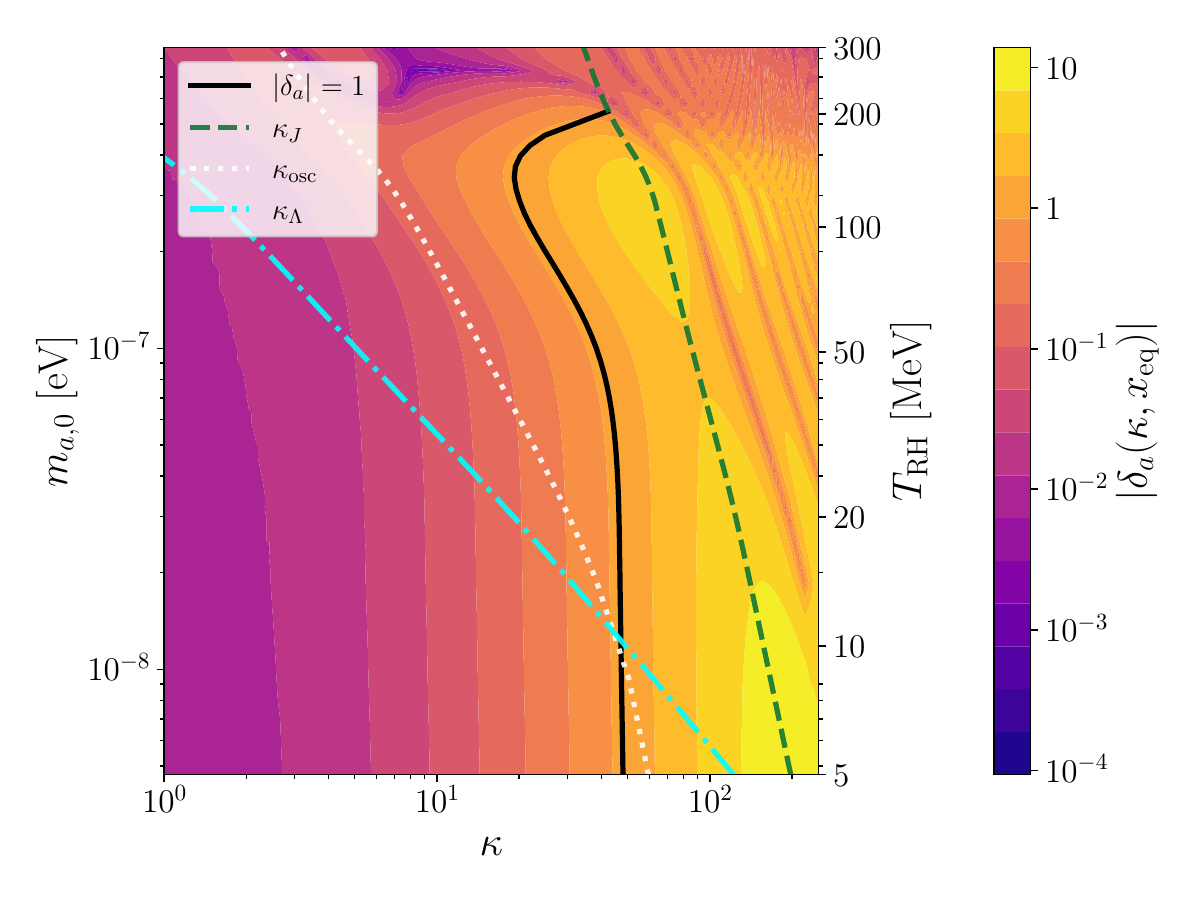}
    \caption{Absolute value of the QCD axion overdensity $\delta_a$ as a function of the comoving wavenumber $\kappa$ and the zero-temperature axion mass $\maz$, for $\Tl = 150$~MeV, and $\thini=1$, evaluated at the MRE. The green dashed lines indicate the Jeans scale $\kappa_J$ at equality, while the black dashed line denotes where the spectrum becomes nonlinear.}
    \label{fig:Mass_spectrum_qcd}
\end{figure}
In Fig.~\ref{fig:Mass_spectrum_qcd} we show the QCD axion overdensity $\delta_a$ as a function of $\maz$ (or, equivalently, $\TRH$, cf.~Eq.~\eqref{relicdensityconstraint}) and the reduced wavenumber $\kappa = k / \kRH$, evaluated at the MRE scale factor $x_{\rm eq}$. The thick black line marks the nonlinearity threshold $|\delta_a| = 1$, while the green dashed line indicates the Jeans threshold of Eq.~\eqref{eq:Jeans}. Two maxima emerge: a global maximum around $\maz \sim 5\times 10^{-9}~$eV corresponding to the lowest cosmologically allowed reheating temperatures $\TRH \simeq \Tbbn$ and a local maximum around $\maz\sim 3\times 10^{-7}~$eV ($\TRH \sim 100$~MeV).

The global maximum corresponds to axion perturbations sourced by the density fluctuations of $\phi$, when axions fall in its gravitational potential. The competition between this gravitational growth and the axion quantum pressure produces a peak in the axion power spectrum at a scale that coincides with the axion quantum Jeans scale at MRE. It leads to the formation of axion stars, as recently explored in Ref.~\cite{Hardy:2026mkp}. In our framework, this corresponds to the gravitational source term $S_2$. In contrast, the local maximum corresponds to a novel effect explored in this work: the temperature dependence of the axion mass couples radiation overdensities to $\delta_a$, producing a source-driven enhancement when $\TRH$ lies just below $\Tl$. The nonlinearity starts at smaller comoving wavenumbers $\kappa \sim 30$, which leaves distinctive imprints on subsequent structure formation. In the next subsection, we explore further the consequences on axion minicluster formation.

\section{Miniclusters}
\label{sec:mini}
The growth of axion density perturbations during the EMD era ceases around the end of reheating, when the field $\phi$ driving the matter-dominated phase decays. Fluctuations in both the radiation and axion DM fluids are released from the deep gravitational potential wells generated on small scales and begin to free-stream~\cite{Visinelli:2018wza}. Thereafter, axion density perturbations evolve as standard non-relativistic matter fluctuations in a radiation-dominated Universe, growing at most logarithmically with the scale factor. Details of the numerical integration and the non-relativistic WKB approximation used to track this evolution up to MRE are provided in Appendix~\ref{sec:WKB}.

Gravitational collapse becomes efficient only once DM dominates the cosmic energy density. Consequently, the $\mathcal{O}(1)$ density contrasts generated during EMD can collapse only around the MRE, defined by $\rho_r = \rho_a$. Assuming axions constitute all the cold DM, equality occurs in the redshift $z_{\rm eq} \simeq 3365$, corresponding to a temperature $T_{\rm eq} \simeq 0.79$~eV~\cite{Planck:2018vyg}. Regions with density contrast $\delta_a$ enter local matter domination earlier, at a redshift $z_{\delta_a}$ given by $(1 + z_{\delta_a}) \simeq (1 + \delta_a)\, (1+ z_{\rm eq})$, and subsequently grow linearly with the scale factor until they become nonlinear and collapse under their gravity~\cite{Kolb:1994fi}. Perturbations with $\delta_a \sim 1$ are therefore expected to collapse shortly after MRE, giving rise to the first generation of dense axion miniclusters~\cite{Kolb:1993zz}.

The density fluctuation spectrum evolved to MRE for the QCD axion is shown in Fig.~\ref{fig:Mass_spectrum_qcd} for several reheating temperatures. The largest fluctuations occur on the smallest scales, down to the Jeans scale, below which quantum pressure suppresses gravitational collapse~\cite{Khlopov:1985fch}. This naturally leads to a hierarchical picture of structure formation. Modes well below the nonlinear scale remain supported by quantum pressure and oscillate, while the first gravitational collapses occur on scales just above the Jeans cutoff. These initial collapses are expected to produce the first generation of axion stars~\cite{Gorghetto:2024vnp}, solitons where gravitational pull is counteracted by quantum pressure; see Refs.~\cite{Ruffini:1969qy, Eby:2019ntd}.

Structure formation then proceeds hierarchically toward larger scales. Modes with $\delta_a \sim 1$ collapse around the MRE, forming abundant DM halos that can be identified with axion miniclusters. Unlike the standard picture of smooth minicluster collapse, these objects are expected to possess a rich internal substructure inherited from the earlier formation of axion stars. 
On still larger scales, clusters of miniclusters form from perturbations with $\delta_a < 1$. Because these fluctuations require additional time to reach nonlinearity, the resulting halos are less dense and therefore are more susceptible to tidal disruption by stellar encounters in the Milky Way~\cite{Tinyakov:2015cgg, Dokuchaev:2017psd, Kavanagh:2020gcy, Shen:2022ltx, OHare:2023rtm}.

Even without dedicated numerical simulations, linear analysis already identifies the characteristic scales associated with this hierarchy of structures. The characteristic minicluster scale is determined by the smallest perturbation mode that becomes nonlinear at MRE. Denoting by $k_{\rm nl}$ the comoving wavenumber satisfying
\begin{equation}
    \delta_a(k_{\rm nl},R_{\rm eq}) = 1\,,
\end{equation}
where $R_{\rm eq}$ is the scale factor at equality; the corresponding physical radius is
\begin{equation}
    r_{\rm nl} \equiv \frac{\pi}{2}\, \frac{R_{\rm eq}}{k_{\rm nl}}\,,
\end{equation}
given that the diameter is the half-wavelength scale $\pi\, R_{\rm eq}/k_{\rm nl}$.

\begin{figure}[t!]
    \centering
    \includegraphics[width=0.7\linewidth]{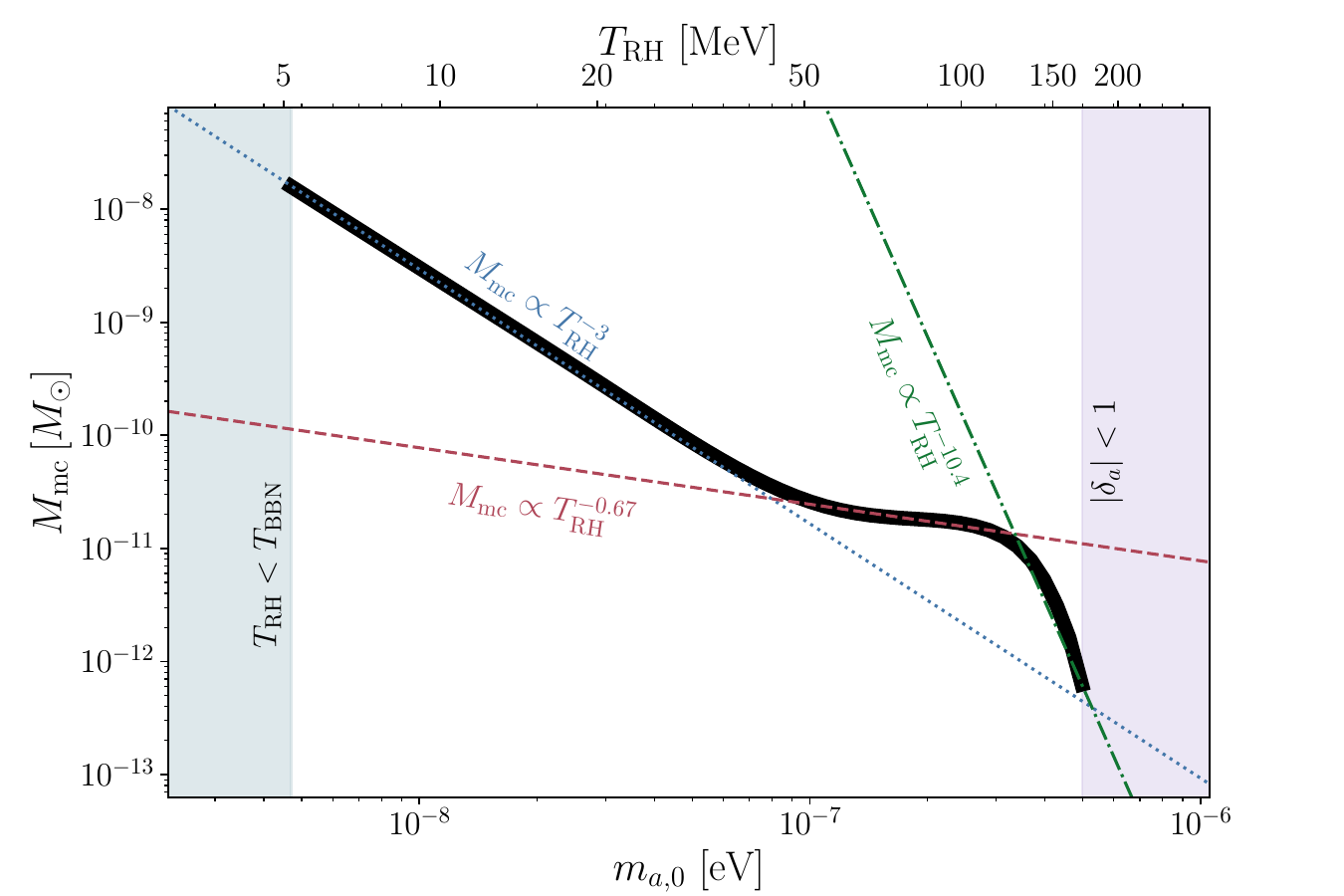}
    \caption{Minicluster mass as a function of the zero-temperature QCD axion mass $\maz$ and the reheating temperature $\TRH$. The solid blue curve shows the numerical result. The dotted red, dashed green, and dash-dotted purple curves represent power-law fits to the numerical solution in different reheating-temperature regimes. The vertical orange dashed line indicates the QCD scale, $\Tl \simeq 150$~MeV.}
    \label{fig:Minicluster_qcd}
\end{figure}
The characteristic minicluster mass $M_{\rm mc}$ is the mass enclosed within a region of radius $r_{\rm nl}$
\begin{align}
    M_{\rm mc} &\simeq \frac{4\pi}{3}\, \rho_a(R_{\rm eq}) \left[1 + \delta_a(k_{\rm nl}, R_{\rm eq})\right] r_{\rm nl}^3 \nonumber\\
    &\simeq 10^{-11} M_\odot \times
    \begin{dcases}
        0.06 \left(\frac{\TRH}{\TQCD}\right)^{-3} &\text{for } \Tbbn \lesssim \TRH \lesssim 35\,\mathrm{MeV}\,, \\
        1.20 \left(\frac{\TRH}{\TQCD}\right)^{-0.67} &\text{for } 35\,\mathrm{MeV} \lesssim \TRH \lesssim \TQCD\,, \\
        0.66 \left(\frac{\TRH}{\TQCD}\right)^{-10.4} &\text{for } \TQCD \lesssim \TRH \lesssim 180\,\mathrm{MeV}\,,
    \end{dcases} 
\end{align}
where the last part corresponds to a numerical fit and is shown in Fig.~\ref{fig:Minicluster_qcd}. At low reheating temperatures, the minicluster mass follows the characteristic scaling $M_{\rm mc} \propto \TRH^{-3}$. This behavior arises because $\kappa_{\rm nl} \simeq 50$ is asymptotically independent of the reheating temperature for $\TRH \ll \Tl$ and $\kRH = H_{\rm RH}\, \RRH \propto \TRH$ up to $\gs$, $\gss$ factors. As reheating approaches the QCD crossover, $\TRH \sim \Tl$, enhanced density fluctuations lead to a decrease in $\kappa_{\rm nl}$, shifting the nonlinear scale toward larger physical radii and producing substantially heavier miniclusters. Simultaneously, the Jeans scale decreases, suggesting the formation of more massive axion stars.

Axion miniclusters and their axion stars will continue evolving through structure formation by accretion, tidal forces, mergers, and wave-collapse~\cite{Zurek:2006sy, Levkov:2018kau, Ellis:2020gtq, Ellis:2022grh, Maseizik:2024uln, Maseizik:2024qya, DSouza:2024uud, Wang:2025vhy, Visinelli:2026bvy}. The broad picture has been outlined in several references and some limited results have been obtained by numerical simulations~\cite{Eggemeier:2019khm, Eggemeier:2019jsu, Dmitriev:2023ipv, Eggemeier:2024fzs}. Axion miniclusters will develop solitonic cores (central axion stars) and a Navarro-Frenk-White profile~\cite{Eggemeier:2019khm} characteristic of DM structure formation. The current halo-mass function is still uncertain, but references agree that the most severe threat to miniclusters and their axion stars is tidal disruption with stars while orbiting galaxies like the Milky Way. 

The Earth-minicluster encounter rate is given by the standard expression for the collision probability in a gas of particles,
\begin{equation}
    \Gamma_{\rm enc} = n_{\rm mc,local}\,\sigma\,v_{\rm rel}\,,
    \label{eq:encounter_rate}
\end{equation}
where $\sigma = \pi R_\text{mc}^2$ is the geometric cross section associated with a minicluster of characteristic radius $R_\text{mc}$, and $v_{\rm rel} \simeq 220~{\rm km/s}$ is the typical relative velocity between the Earth and dark matter in the Galactic halo.

The virialized radius of the minicluster $R_\text{mc}$ is obtained from its mass and internal density,
\begin{equation}
    R_\text{mc} = \left(\frac{3 M_\text{mc}}{4\pi \rho_{\rm mc}}\right)^{1/3}\,,
\end{equation}
where $\rho_{\rm mc}$ denotes the average density within the virialized object. To estimate $\rho_{\rm mc}$, we follow the spherical collapse analysis of Ref.~\cite{Kolb:1994fi}, which is obtained for large-amplitude isocurvature fluctuations that become nonlinear before matter--radiation equality, but we apply it to our scenario, assuming the spherical collapse dynamics is independent of the perturbation origin. In that framework, the final virialized core density of a clump seeded by an initial overdensity $\xi \equiv \delta\rho_{\rm DM}/\rho_{\rm DM}$ is~\cite{Kolb:1994fi, Tinyakov:2015cgg} 
\begin{equation}
    \rho_{\rm mc} \simeq 140\,\xi^{3}(\xi+1)\,\rho_{\rm EQ}\sim 7\times 10^6 \,\xi^3\,(\xi+1)\,\mbox{GeV/cm}^3\,,
    \label{eq:rho_mc}
\end{equation}
where $\rho_{\rm EQ}$ is the matter energy density at matter--radiation equality. We use $\xi\sim 1$ since our estimates are based on a linear analysis. Then, the encounter rate Eq.~\eqref{eq:encounter_rate} is given by

\begin{align}
        \Gamma_{\rm enc} &= \pi\,v_{\rm rel}\,\rho_{\rm DM,local}\left(\frac{3}{4\pi\,\rho_{\rm mc}}\right)^{\!2/3} M_{\rm mc}^{-1/3},\nonumber\\
    &\simeq 2 \times 10^{-6}~\text{year}^{-1} \times
    \begin{dcases}
        2.5 \left(\frac{\TRH}{\TQCD}\right) &\text{for } \Tbbn \lesssim \TRH \lesssim 35\,\mathrm{MeV}\,, \\
        0.91 \left(\frac{\TRH}{\TQCD}\right)^{0.22} &\text{for } 35\,\mathrm{MeV} \lesssim \TRH \lesssim \TQCD\,, \\
        1.12 \left(\frac{\TRH}{\TQCD}\right)^{3.5} &\text{for } \TQCD \lesssim \TRH \lesssim 180\,\mathrm{MeV}\,.
    \end{dcases}
\end{align}

Axion miniclusters have a rich phenomenology, which potentially enables their indirect discovery but also hampers the direct detection of DM. Simulations of post-inflationary axion scenarios support the expectation that an $\mathcal{O}(1)$ fraction of DM can be bound into dense miniclusters when the primordial density fluctuations reach order unity~\cite{Zurek:2006sy, Eggemeier:2019jsu, Eggemeier:2024fzs, Gorghetto:2024vnp, Hardy:2026mkp}. The residual density in voids then sets a lower bound on the smooth axion component available to terrestrial haloscope searches~\cite{Eggemeier:2022hqa}. However, recent estimates of the disruption probability indicate that even if axion miniclusters and stars are too dense to be disrupted, the next generations of miniclusters, i.e. the clusters of miniclusters will be disrupted, releasing the trapped axions into tidal streams that overlap and restore the local DM density around the Earth to $\sim 80\%$ of its standard value~\cite{OHare:2023rtm}. The surviving miniclusters and stars can be detected by a number of probes, such as microlensing~\cite{Fairbairn:2017sil, Schiappacasse:2021zlr, Ellis:2022grh, Prabhu:2020pzm}, femtolensing~\cite{Kolb:1995bu, Katz:2018zrn}, see also Refs.~\cite{Dai:2019lud, Muller:2024pwn} and radio emission~\cite{Iwazaki:2014wka, Raby:2016deh, Eby:2017xaw, Edwards:2020afl, Kavanagh:2020gcy, Amin:2020vja, Amin:2021tnq, Eby:2021ece, Fox:2023xgx, Visinelli:2024tyw, Walters:2024vaw, Maseizik:2024uln, Fox:2025tqa}.

\section{Conclusions}
\label{sec:conclusions}
In this work, we have studied the growth of axion dark-matter perturbations during an early matter-dominated era. We focused on axions produced through the standard misalignment mechanism in a pre-inflationary PQ-breaking scenario, allowing for a temperature-dependent mass as motivated by the QCD axion and by ALPs whose mass is generated by a hidden confining sector.

The key effect identified in this work is that fluctuations in the radiation temperature induce fluctuations in the axion mass. This generates an additional source term in the perturbed Klein--Gordon equation, directly coupling radiation overdensities to axion perturbations. During EMD, radiation perturbations sourced by the decay of the matter-like field can become much larger than in standard radiation domination. When the axion mass is still temperature dependent, these enhanced radiation perturbations are efficiently transferred to the axion sector, leading to a significant amplification of small-scale axion overdensities.

For generic ALPs with $m_a(T)=m_{a,0}(T_\Lambda/T)^b$ above the saturation temperature $T_\Lambda$, we found that the enhancement is largest when axion oscillations begin during EMD and the mass saturates close to reheating. Once the mass reaches its zero-temperature value, the temperature-induced source shuts off and axion pressure suppresses sufficiently small scales. 

We also applied the framework to the QCD axion, using the lattice-motivated temperature dependence of the mass. The strongest enhancement occurs for reheating temperatures below but close to the QCD scale. In the corresponding region of parameter space, axion overdensities can become nonlinear by matter--radiation equality, providing a mechanism for forming axion miniclusters even in a pre-inflationary PQ-breaking scenario, without relying on topological defects or primordial axion isocurvature fluctuations.

The small-scale structures produced in this way can include axion miniclusters and, near the Jeans scale, dense axion-star-like cores. Their detailed properties require nonlinear simulations, including gravitational collapse, wave dynamics, mergers, accretion, and tidal disruption in galactic environments. More broadly, our results show that temperature-dependent axion masses provide a sensitive probe of the pre-BBN expansion history, and that small-scale axion structure can carry information about both axion microphysics and low-temperature reheating.

\section*{Acknowledgments}
AA and PA acknowledge support from FONDECYT project No.~1251613.  This publication is based on work from the COST Action ``COSMIC WISPers'' (CA21106). NB thanks the Universidad Técnica Federico Santa María for its hospitality.

\appendix
\addappheadtotoc
\section{Perturbed fluid equations} \label{A_perts}
In this work, we treat the scalar field $\phi$ and the SM radiation $r$ as perfect fluids, adopting a similar treatment to that of Refs.~\cite{Erickcek:2011us, Fan:2014zua, Marcondes:2016reb}. Each fluid $\sigma$ is characterized by an energy density $\rho_\sigma$, a pressure $p_\sigma$, and a stress-energy tensor
\begin{equation}
    {T^{(\sigma)}}^\mu_\nu = \left(\rho_\sigma + p_\sigma\right) u^\mu u_\nu + \delta^\mu_\nu\, p_\sigma\,,
    \label{perfectTmunu}
\end{equation}
with $\sigma = \phi$, $r$. At the background level, we consider a Universe described by the FLRW metric
\begin{equation}
    ds^2 = -dt^2 + R^2(t)\,\delta_{ij}\,dx^i dx^j \, .
\end{equation}
We define the four-velocity in the rest frame of each fluid by the normalization condition $u_\mu u^\mu = -1$, such that $u^\mu = (1,0,0,0)$. In our setup, the scalar field and radiation interact
\begin{equation} \label{eq:Conservation_A}
    \nabla_\mu\, {T^{(\sigma)}}^\mu_\nu =  Q^{(\sigma)}_\nu,
\end{equation}
so the stress tensor of each component is not conserved. However, the sum over all species must satisfy the covariant conservation law of the total energy–momentum tensor
\begin{equation}
    \sum_{\sigma=\phi,r} \nabla_\mu\, {T^{(\sigma)}}^\mu_\nu = 0\,,
\end{equation}
where
\begin{align}
    Q^{(\phi)}_\nu &= \Gphi\, T^{(\phi)}_{\mu\nu}\, u^\mu\,, \label{eq:Injection_entropy}\\ 
    Q^{(r)}_\nu &= -Q^{(\phi)}_\nu
\end{align}
represent the energy injected through decays of $\phi$ into SM radiation~\cite{Erickcek:2011us}. It follows from Eq.~\eqref{eq:Conservation_A} that the background continuity equation for each fluid component is
\begin{equation}
    \dot{\rho}_{\sigma} + 3\, H \rho_\sigma\, (1 + w_\sigma) = - Q_0^{(\sigma)}, \label{eq:continuity_back}
\end{equation}
where $w_\sigma$ is the equation-of-state parameter of $\sigma$.

The perturbed metric in the Newtonian gauge is given by 
\begin{equation}
    ds^2 = - (1 + 2\Phi)\, dt^2 + R^2(t)\, (1 - 2\Psi)\, \delta_{ij}\, dx^i dx^j\,,
    \label{eq:metric_newtonian_A}
\end{equation}
where $\Phi$ and $\Psi$ are the Newtonian potential and the spatial curvature perturbation, respectively~\cite{Dodelson:2003ft}. From now on, we will neglect the anisotropic stress, setting $\Phi = \Psi$. The energy-momentum tensor can be decomposed linearly into a background ($\overline{T}^{(\sigma)}$) and a perturbed ($\delta {T^{(\sigma)}}$) contribution as
\begin{equation}
    {T^{(\sigma)}}^\mu_\nu = {\overline{T}^{(\sigma)}}^\mu_\nu + \delta {T^{(\sigma)}}^\mu_\nu\,.
\end{equation}
In the same way, the energy density and pressure can be decomposed as
\begin{align}
    \rho_{\sigma} &= \bar{\rho}_{\sigma} + \delta\rho_{\sigma}, \\
    p_{\sigma} &= \bar{p}_{\sigma} + \delta p_{\sigma}.
\end{align}
For the fluids considered here, we take
\begin{equation}
    w_\phi = c_\phi^2 = 0
    \qquad \text{and} \qquad
    w_r = c_r^2 = \frac13\,,
\end{equation}
where $c_\sigma^2 \equiv \delta p_\sigma/\delta\rho_\sigma$ for the barotropic fluids used in this work. For each fluid, we can write the individual components of the perturbed stress-energy tensor as
\begin{align}
    \delta {T^{(\sigma)}}^0_0 &= -\delta\rho_{\sigma},  \label{Tensors_perturbed1}\\
    \delta {T^{(\sigma)}}^0_i &= \rho_{\sigma}\big(1+w_{\sigma}\big)\partial_i v_{\sigma}, \\
    \delta {T^{(\sigma)}}^i_0 &= -\rho_{\sigma}\big(1+w_{\sigma}\big)g^{ij}\partial_j v_{\sigma}, \\
    \delta {T^{(\sigma)}}^i_j &= c_{\sigma}^{2} \delta\rho_{\sigma} \delta^{i}_{j}, \label{Tensors_perturbed4}
\end{align}
where we use the condition $u^\mu u_\mu = -1$ to write the perturbed four-velocity as $u_\mu = (-1 -\Phi,\, \partial_i v_{\sigma})$, with $v_{\sigma}$ being the velocity potential of the fluid $\sigma$.

The energy transfer terms $Q^{(\sigma)}_\nu$ can be decomposed into a background and a perturbed contribution as
\begin{equation}
    Q_\nu^{(\sigma)} = \overline{Q}_\nu^{(\sigma)} + \delta Q_\nu^{(\sigma)};
\end{equation}
with this, the covariant conservation equation for the perturbed energy–momentum tensor of each fluid reads
\begin{equation}
    \nabla_\mu \,\delta {T^{(\sigma)}}^\mu_\nu + \delta\Gamma^\mu_{\mu\alpha}\, {\overline{T}^{(\sigma)}}^\alpha_\nu - \delta\Gamma^\alpha_{\mu\nu}\, {\overline{T}^{(\sigma)}}^\mu_\alpha = \delta Q_\nu^{(\sigma)},
    \label{eq:conservation_tensor_perturbed}
\end{equation}
where $\delta \Gamma^{\lambda}_{\mu\nu}$ denotes the perturbation of the Christoffel symbols. For our metric, the non-vanishing components are
\begin{align}
    \delta\Gamma^0_{00} &= \dot{\Phi}, & 
    \delta\Gamma^i_{00} &= \frac{1}{R^2}\partial_i\Phi, \nonumber \\
    \delta\Gamma^0_{0i} &= \partial_i\Phi, & 
    \delta\Gamma^i_{0j} &= -\dot{\Phi}\delta^i_j, \\
    \delta\Gamma^0_{ij} &= -R^2\big(4H\Phi + \dot{\Phi}\big)\delta_{ij}, \qquad & 
    \delta\Gamma^i_{jk} &= -\partial_j\Phi\delta^i_k - \partial_k\Phi\delta^i_j + \partial_i\Phi\delta_{jk}. \nonumber
    \label{eq:christoffel_perturbed}
\end{align}
As commonly used, we move to new variables, the overdensity $\delta_\sigma$ and the divergence of the conformal velocity $\theta_\sigma$, defined as
\begin{equation}
    \delta_\sigma \equiv \frac{\delta\rho_\sigma}{\rho_\sigma}, \qquad \theta_\sigma \equiv \frac{1}{R}\nabla^2 v_\sigma.
\end{equation}
We obtain the evolution equations for density and velocity perturbations by applying the perturbed conservation law of the energy–momentum tensor, Eq.~\eqref{eq:conservation_tensor_perturbed}, to the temporal and spatial components, and by including on the right-hand side the corresponding energy transfer term $Q_\nu^{(\sigma)}$, following 
\begin{align}
    \dot{\delta}_{\sigma} + 3H\left(c^2_{\sigma} - w_{\sigma}\right)\delta_{\sigma} + (1 + w_{\sigma})\left( \frac{\theta_{\sigma}}{R} - 3\dot{\Phi} \right) &= \frac{1}{\rho_{\sigma}}\left[\overline{Q}_0^{(\sigma)}\delta_{\sigma}- \delta Q_0^{(\sigma)}\right], \label{eq:delta_sigma_Q}\\
    \dot{\theta}_{\sigma} + \left[H(1-3\,w_\sigma)\right]\theta_{\sigma} + \frac{c^2_{\sigma}}{1 + w_{\sigma}} \frac{\nabla^2 \delta_{\sigma}}{R} + \frac{\nabla^2 \Phi}{R} &= \frac{1}{\rho_{\sigma}}\left[ \frac{\partial_i\delta Q_i^{(\sigma)}}{R(1+w_{\sigma})}+\overline{Q}_0^{(\sigma)}\theta_{\sigma}\right]. \label{eq:theta_sigma_Q} 
\end{align}
Expanding the source term in Eq.~\eqref{eq:Injection_entropy} up to first order gives
 \begin{align} 
    \delta Q_0^{(\phi)} &= +\Gphi\, \rp (\delta_\phi + \Phi)\,, \label{Q_0_c_2A} \\
    \delta Q_i^{(\phi)} &= -\Gphi\, \rp \,\partial_i v_\phi\,. \label{Q_i_c_2A}
\end{align} 
Using Eqs.~\eqref{Q_0_c_2A} and~\eqref{Q_i_c_2A} in Eqs.~\eqref{eq:delta_sigma_Q} and~\eqref{eq:theta_sigma_Q}, and changing to Fourier space, by taking $\vec \nabla \to i\, \vec{k}$, we obtain the evolution equations for the density and velocity perturbations of the scalar field and radiation as
\begin{align}
    \dot{\delta}_{\phi} &= -{\Gamma}_{\phi}\Phi - \frac{\theta_{\phi}}{R} + 3\dot{\Phi}, \label{eq:deltaphiA}\\
    \dot{\theta}_{\phi} &= -H\theta_{\phi} + \frac{k^{2}\Phi}{R}, \label{eq:thetaphiA}\\
    \dot{\delta}_{r} &= \frac{{\rho}_{\phi}{\Gamma}_{\phi}}{\rho_{r}} \left( \delta_{\phi} - \delta_{r} + \Phi  \right) - \frac{4}{3}\frac{\theta_{r}}{R} + 4\dot{\Phi}, \label{eq:deltarA}\\
    \dot{\theta}_{r} &= \frac{{\rho}_{\phi}\,{\Gamma}_{\phi}}{\rho_{r}} \left( \frac{3}{4}\theta_{\phi} - \theta_{r} \right) + \frac{k^{2}}{R}\left(\frac{\delta_{r}}{4}+\Phi \right). \label{eq:thetarA}
\end{align}
Furthermore, the evolution of the gravitational potential can be derived from the $\mu = \nu = 0$ component of the perturbed Einstein equations, relating the density perturbations to the metric perturbations, and takes the form
\begin{equation}
    3H(\dot{\Phi} + H\Phi) + \frac{k^2}{R^2}\Phi = -\frac{1}{2M_{\text{Pl}}^2}\sum_\sigma \delta\rho_\sigma .
    \label{eq:poisson_pert_AP}
\end{equation}
This is the full Poisson equation, which describes how the density perturbations of all components source the metric perturbations. In this expression, the term $\sum_\sigma \delta\rho_\sigma$ denotes the total density perturbation obtained by summing all the components of the fluid. For convenience, we rewrite Eqs.~\eqref{eq:deltaphiA} to~\eqref{eq:poisson_pert_AP} in terms of the scale factor, which yields
\begin{align}
    \delta'_\phi &= -\frac{\Gphi}{R H} \Phi - \frac{1}{R^2 H} \theta_\phi + 3\Phi' ,\label{eq:deltaphiAR}\\
    \delta'_r &= \frac{\rp}{\rR} \frac{\Gphi}{R H} (\delta_\phi - \delta_r + \Phi ) - \frac{4}{3} \frac{1}{R^2 H} \theta_r + 4\Phi', \label{eq:deltarAR}\\
    \theta'_\phi &= \frac{k^2}{R^2 H} \Phi - \frac{1}{R} \theta_\phi ,\label{eq:thetaphiAR}\\
    \theta'_r &= \frac{\rp}{\rR} \frac{\Gphi}{R H} \left( \frac{3}{4} \theta_\phi - \theta_r \right) + \frac{k^2}{R^2 H} \left( \frac{\delta_r}{4} + \Phi \right), \label{eq:thetarAR}\\
    \Phi' &= -\left[ \frac{1}{6 M_{\rm Pl}^2 R H^2} \left(\rp \delta_\phi + \rR \delta_r + \rho_a \delta_a \right) + \left( \frac{k^2}{3 R^3 H^2} + \frac{1}{R} \right) \Phi \right] .\label{eq:phiAR}
\end{align}
The axion overdensity appears in Eq.~\eqref{eq:phiAR}, but can be safely neglected since $\rho_a \ll \rR, \rp$.

\subsection*{Initial conditions} \label{Initial-conditions-NSC}
Now we derive the initial conditions for the system of Eqs.~\eqref{eq:deltaphiAR} to~\eqref{eq:phiAR}, following Refs.~\cite{Erickcek:2011us, Blinov:2019jqc}. We assume that all relevant modes are initially superhorizon ($k \ll H R$). In this regime, perturbations are constant in the Newtonian gauge, so $\delta'_\sigma = 0$. During the EMD era, the gravitational potential remains approximately constant
\begin{equation}
    \Phi(R) \simeq \Phi(\Rini) \equiv \Phi_{\rm p},
\end{equation}
where $\Phi_{\rm p}$ denotes the initial Newtonian-potential amplitude. In the numerical analysis, we take $\Phi_{\rm p} = \sqrt{\mathcal{A}_s}$ as a convenient normalization, with $\mathcal{A}_s \simeq 2.101 \times 10^{-9}$ the amplitude of the primordial scalar power spectrum measured by Planck~\cite{Planck:2018vyg}.

Starting from Eq.~\eqref{eq:phiAR}, we neglect terms of order $\mathcal{O}(k^2)$ in the superhorizon limit. Since the Universe is $\phi$-dominated, we use $3\, M_{\rm Pl}^2\, H^2 \simeq \rho_\phi$ and then
\begin{equation}
    \frac{1}{6\, M_{\rm Pl}^2\, H^2} \left(\rho_\phi \delta_\phi + \rho_r \delta_r + \rho_a \delta_a \right) + \Phi \simeq \frac12 \left(\delta_\phi + \frac{\rho_r}{\rho_\phi}\,\delta_r + \frac{\rho_a}{\rho_\phi}\,\delta_a \right) + \Phi \simeq 0\,.
\end{equation}
Since $\rho_\phi \gg \rho_r, \rho_a$, we neglect the contributions from $\delta_r$ and $\delta_a$ to obtain
\begin{equation}
    \delta_\phi(\Rini) = -2\,\Phi_{\rm p}.
    \label{in_delta_phi}
\end{equation}
Next, we study the radiation overdensity in Eq.~\eqref{eq:deltarAR}. We use $\Phi' \simeq 0$ and the superhorizon condition $\delta_r' \simeq 0$. From Eq.~\eqref{eq:fluid_perturbed_main}, $\theta_\sigma \equiv \nabla^2 v_\sigma/R$, so $\theta_r \sim \mathcal{O}(k^2)$ and can be neglected. With these approximations, we obtain
\begin{equation}
    \delta_\phi(\Rini) - \delta_r(\Rini) + \Phi_{\rm p} = 0.
    \label{eq:delta_relation_ini}
\end{equation}
Using Eq.~\eqref{in_delta_phi}, we find
\begin{equation}
    \delta_r(\Rini) = -\Phi_{\rm p}.
\end{equation}
Additionally, for the initial condition of the velocity divergence, we set
\begin{equation}
    \theta_\phi(\Rini) = \theta_r(\Rini) = 0.
\end{equation}
This is justified by the fact that superhorizon modes remain frozen and do not develop spatial gradients, so the velocity divergence is negligible. Furthermore, as shown by the analytical approximation during the EMD era (see Eq.~\eqref{eq:thetaphi_an}), the initial value term becomes progressively irrelevant as the scale factor evolves and remains subleading compared to the other terms of $\theta_\sigma(R)$.

\section{Analytical solutions} \label{sec:analytics}
In the following, analytical solutions for the evolution of the background and its perturbations are presented, valid deep in the EMD era. Consequently, we assume that $\rp \gg \rR$, $H \gg \Gphi$, and a negligible initial SM radiation density.

\subsection*{Background}
The energy densities for $\phi$ and the SM radiation in Eqs.~\eqref{eq:boltzmann1_rew} and~\eqref{eq:boltzmann2_rew} scale as
\begin{align}
    \rp(R) &\simeq \rp(\Rini) \left(\frac{\Rini}{R}\right)^3, \label{eq:densities_an1}\\
    \rR(R) &\simeq \frac{2\sqrt{3}}{5}\, M_\text{Pl}\, \Gphi\, \sqrt{\rp(\Rini)} \left(\frac{\Rini}{R}\right)^{3/2},
    \label{eq:densities_an2}
\end{align}
while the Hubble expansion rate scales as
\begin{equation}
    H(R) \simeq  H(\Rini)\left(\frac{\Rini}{R}\right)^{3/2},
    \label{eq:Hubble_R_in}
\end{equation}
with
\begin{equation}
    H^2(\Rini) \equiv \frac{\rho_\phi(\Rini)}{3\, M_{\rm Pl}^2}\,.
\end{equation}

\subsection*{Fluid perturbations during EMD}
As the gravitational potential remains constant during the EMD era with a value $\Phi(R) \simeq \Phi_{\rm p}$, Eq.~\eqref{eq:phi} becomes
\begin{equation}
    \Phi' \simeq 0\,.
\end{equation}
Then, the derivative of the velocity divergence for $\phi$ in Eq.~\eqref{eq:thetaphi} can be solved, giving rise to
\begin{equation}
    \theta_\phi(R) \simeq \theta_{\phi,{\rm i}}\frac{\Rini}{R}+ \frac23\, \frac{k^2\, \Phi_{\rm p}}{\kini} \left[\sqrt{\frac{R}{\Rini}}- \frac{\Rini}{R}\right] \simeq \frac23\, \frac{k^2\, \Phi_{\rm p}}{\kini} \sqrt{\frac{R}{\Rini}},
    \label{eq:thetaphi_an}
\end{equation}
with $\kini \equiv \Rini\, H(\Rini)$. We use this approximate solution in Eq.~\eqref{eq:deltaphiAR} and solve for $\delta_\phi$, obtaining
\begin{align}
    \delta_\phi(R) &\simeq \delta_{\phi,{\rm i}} - \frac23\, \Phi_{\rm p} \left(\frac{k}{\kini}\right)^2 \frac{R}{\Rini} \left[1 + 2 \left(\frac{\Rini}{R}\right)^{3/2} - 3\, \frac{\Rini}{R}\right] \nonumber\\
    &\simeq -2\, \Phi_{\rm p} - \frac{2}{3}\, \Phi_{\rm p} \left(\frac{k}{\kini}\right)^2 \frac{R}{\Rini} = -2\, \Phi_{\rm p} \left[1 + \frac{\kappa^2}{3}\, \frac{R}{\RRH}\right].
\end{align}
It is interesting to note that during the EMD era, nonlinearity $|\delta_\phi| = 1$ can only be reached for modes corresponding to $\kappa \gtrsim \sqrt{3/(2\, \Phi_\text{p})} \simeq 181$, at $R = \RRH$.

The common source term in Eqs.~\eqref{eq:deltarAR} and~\eqref{eq:thetarAR} can be approximated using the analytical background solutions given in Eqs.~\eqref{eq:densities_an1} to~\eqref{eq:Hubble_R_in}. We obtain $\frac{\rho_\phi}{\rho_r}\,\frac{\Gphi}{H} \simeq \frac52$. Using this result, together with the analytical solutions for $\delta_\phi$ and $\theta_\phi$, the system is reduced to
\begin{align}
    \delta'_r &\simeq -\frac{5}{2R}\,\delta_r- \frac{4}{3\, \kini\, \sqrt{\Rini R}}\,\theta_r- \frac{5}{2R}\,\Phi_{\rm p}\left[ 1 + \frac{2}{3}\left(\frac{k}{\kini}\right)^2 \frac{R}{\Rini} \right],
    \label{eq:deltar_edo_an} \\
    \theta'_r &\simeq \frac{k^2}{4\, \kini\, \sqrt{\Rini R}}\,\delta_r- \frac{5}{2R}\,\theta_r+ \frac{9}{4}\,\frac{k^2 \Phi_{\rm p}}{\kini\, \sqrt{\Rini R}}\,.
    \label{eq:thetar_edo_an}
\end{align}
Next, we differentiate Eq.~\eqref{eq:deltar_edo_an} with respect to $R$, and then substitute $\theta'_r$ and $\theta_r$ using Eqs.~\eqref{eq:thetar_edo_an} and~\eqref{eq:deltar_edo_an}, respectively, to obtain
\begin{equation}
    \delta''_r + \frac{11}{2R} \delta'_r + \frac{1}{R^2} \left[ 5 + \frac{1}{3} \left(\frac{k}{\kini}\right)^2 \frac{R}{\Rini} \right] \delta_r \simeq - \frac{1}{R^2} \Phi_{\rm p} \left[ 5 +8\left(\frac{k}{\kini}\right)^2 \frac{R}{\Rini} \right].
    \label{eq:delta_r_edo}
\end{equation}
For the subhorizon modes, the terms proportional to $k^2$ dominate over the contributions independent of $k$ within the squared brackets. The maximal growth of the subhorizon perturbation can be found by setting  $\delta''_r = \delta'_r = 0$, and corresponds to
\begin{equation}
    \delta_r \simeq -24\, \Phi_{\rm p} = 24\, \delta_r(\Rini)\,.
\end{equation}
This corresponds to the plateau observed in the lower left panel of Fig.~\ref{fig:delta}. Additionally, it is possible to solve Eq.~\eqref{eq:delta_r_edo} analytically; we find
\begin{align}
    \delta_r(\tilde{x}) &\simeq \Phi_{\rm p} \Bigg[ -24 + \frac{345}{\tilde{\kappa}^2 \tilde{x}} - \frac{3105}{2 \tilde{\kappa}^4 \tilde{x}^2} \nonumber \\
    &\quad + \frac{1}{4 \tilde{\kappa}^5 \tilde{x}^{5/2}} \Big\{ 46 \tilde{\kappa} \left( 135 - 30 \tilde{\kappa}^2 + 2 \tilde{\kappa}^4 \right) \cos \varphi(\tilde{x}) + 115 \sqrt{3} \left( 27 - 18 \tilde{\kappa}^2 + 2 \tilde{\kappa}^4 \right) \sin \varphi(\tilde{x}) \Big\} \Bigg],
    \label{eq:deltar_exact}
\end{align}
where we have used the initial condition $\delta_r'(\Rini)=0$, and the dimensionless variables $\tilde{x} \equiv R/\Rini$ and $\tilde{\kappa} \equiv k/\kini$. The phase is given by
\begin{equation}
    \varphi(\tilde{x}) \equiv \frac{2\, \tilde{\kappa}}{\sqrt{3}} \left( \sqrt{\tilde{x}} - 1 \right).
\end{equation}
Finally, the velocity divergence for the radiation is
\begin{align}
    \theta_r(\tilde{x}) \simeq  \frac{\theta_{r, \rm i}}{\tilde{x}^{5/2}} &+ \frac{\kini\, \Phi_{\rm p}}{16 \tilde{\kappa}^4 \tilde{x}^{5/2}} \Bigg[5 \left( 1863 - 1242 \tilde{\kappa}^2 \tilde{x} + 138 \tilde{\kappa}^4 \tilde{x}^2 - 4 \tilde{\kappa}^6 (\tilde{x}^3 - 1) \right) \nonumber\\ 
    &- 345 (27 - 18 \tilde{\kappa}^2 + 2 \tilde{\kappa}^4) \cos(\varphi(\tilde{x})) + 46\sqrt{3} \tilde{\kappa} (135 - 30 \tilde{\kappa}^2 + 2 \tilde{\kappa}^4) \sin(\varphi(\tilde{x})) \Bigg].
\end{align}

\section{Axion perturbations} \label{B_perts}
It is useful to express the axion field in terms of a dimensionless angular variable, defined as $\theta=a/f_a$, with values within $[-\pi,\pi]$. We decompose the field into a homogeneous component and a first-order perturbation,
\begin{equation}
    \theta(t,\vec{x}) = \theta_0(t) + \delta\theta(t,\vec{x}),
\end{equation}
with
\begin{equation}
    \delta\theta(t,\vec{x}) = \int \frac{d^3k}{(2\pi)^3}\, \delta\theta_k(t)\, e^{i\, \vec{k}\cdot\vec{x}} .
    \label{eq:axion_field1}
\end{equation}
Since the axion mass depends on temperature, the mass term also acts as a source of perturbations. Therefore, the mass term can be expanded to the first order as
\begin{equation}
    \ma^2(T) = \ma^2(\overline{T}) + \delta\ma^2\,, \qquad \text{with} \qquad \delta\ma^2 = \left. \frac{d\ma^2}{dT}\right|_{\overline{T}} \delta T \,,
    \label{eq:mass_exp}
\end{equation}
where we used the first-order expansion of the temperature $T(t,k)= \overline{T}(t) + \delta T(t,k)$, with $\overline{T}$ the background temperature. 

Then, using Eqs.~\eqref{eq:axion_field1} and~\eqref{eq:mass_exp}, we expand at linear order the derivative of the axion potential with respect to $\theta$ as
\begin{align}
   \frac{\partial V(\theta, T)}{\partial \theta} &= \ma^2(T)\, f_a^2\, \sin\theta  \nonumber\\
   &\simeq \ma^2(\overline{T})\, f_a^2\, \sin{\theta_0} + \ma^2(\overline{T})\, f_a^2\, \cos{\theta_0}\, \delta\theta_k + \left. \frac{d\ma^2}{dT} \right|_{\overline{T}}\, f_a^2\, \sin{\theta_0}\, \delta T\,.
   \label{eq:dV_expanded}
\end{align}
Next, we can construct the Klein-Gordon equation 
\begin{equation}
    \frac{1}{\sqrt{-g}}\, \partial_\mu \left(\sqrt{-g}\, g^{\mu\nu}\, \partial_\nu \theta \right) - \frac{1}{f_a^2} \frac{\partial V(\theta, T)}{\partial \theta} = 0\,.
\end{equation}
Using the perturbed metric in Eq.~\eqref{eq:metric_newtonian}, the first term can be written as
\begin{equation}
    \frac{1}{\sqrt{-g}}\, \partial_\mu \left(\sqrt{-g}\, g^{\mu\nu}\, \partial_\nu \theta \right)\simeq -(1 - 2\Phi)\left( \ddot{\theta}_0 + 3H\dot{\theta}_0 \right) - \left( \delta\ddot{\theta}_k + 3H\delta\dot{\theta}_k + \frac{k^2}{R^2}\delta\theta_k \right)+ 4\dot{\Phi}\dot{\theta}_0 \,,
\end{equation}
we obtain the perturbed equation of motion for the axion field
\begin{equation}
    \frac{\partial^2 \delta \theta_k}{\partial t^2} + 3 H \frac{\partial \delta\theta_k}{\partial t} + \frac{k^2}{R^2} \delta \theta_k + \ma^2\, \cos\theta_0\, \delta \theta_k = 4 \dot \Phi\, \dot \theta_0 - 2 \Phi\, \ma^2\, \sin\theta_0 - \frac14 \frac{d \ma^2}{dT} \sin\theta_0\, \overline{T}\, \delta_r,
    \label{eq:perturbation_t_ap}
\end{equation}
where we have used $\delta \rR/\rR = 4 \, \delta T/\overline{T}$, ignoring the variation of the effective relativistic degrees of freedom. The perturbed fluid quantities can be obtained using the energy-momentum tensor of a scalar field
\begin{equation}
    T^\mu_\nu = f_a^2\, g^{\mu\lambda}\, \partial_\nu\theta\, \partial_\lambda\theta - \delta^\mu_\nu \left[ \frac{f_a^2}{2}\, g^{\lambda\rho}\, \partial_\lambda\theta\, \partial_\rho\theta + V(\theta) \right].
\end{equation}
Using the perturbed metric in Eq.~\eqref{eq:metric_newtonian} and expanding the field to first order as in Eq.~\eqref{expantion_field}, we isolate the first-order perturbed temporal and spatial components of the energy-momentum tensor as
\begin{align}
    \delta T^0_0 &= f_a^2 \left[ \Phi\, \dot{\theta}_0^2 - \dot{\theta}_0\, \delta\dot{\theta}_k - \ma^2(\overline{T}) \sin{\theta_0}\, \delta\theta_k - \frac{\overline{T}}{4} \left. \frac{d\ma^2}{dT} \right|_{\overline{T}} (1 - \cos{\theta_0})\, \delta_r \right],\\
    \delta T^i_i &= 3 f_a^2 \left[ \dot{\theta}_0\, \delta\dot{\theta}_k - \Phi\, \dot{\theta}_0^2 - \ma^2(\overline{T}) \sin{\theta_0}\, \delta\theta_k - \frac{\overline{T}}{4} \left. \frac{d\ma^2}{dT} \right|_{\overline{T}} (1 -\cos{\theta_0})\, \delta_r \right].
\end{align}
In the regime $T>\Tl$, the mass of the axion still depends on the temperature. In this case, using Eq.~\eqref{eq:axion_mass}, we can write
\begin{equation}
    \frac{\overline{T}}{4} \left. \frac{d\ma^2}{dT} \right|_{\overline{T}} = -\frac{b}{2}\, \ma^2\, .
\end{equation}
Then, using Eqs.~\eqref{Tensors_perturbed1} and~\eqref{Tensors_perturbed4}, the perturbed axion energy density and pressure can be written as
\begin{align}
    \delta \rho_a &= f_a^2 \left[ \dot{\theta}_0\, \delta\dot{\theta}_k  - \Phi\, \dot{\theta}_0^2 + \ma^2(\overline{T}) \sin{\theta_0}\, \delta\theta_k - \frac{b}{2} \ma^2(\overline{T})\, (1 - \cos{\theta_0})\, \delta_r \right],\\
    \delta p_a &= f_a^2 \left[ \dot{\theta}_0\, \delta\dot{\theta}_k - \Phi\, \dot{\theta}_0^2 - \ma^2(\overline{T}) \sin{\theta_0}\, \delta\theta_k + \frac{b}{2} \ma^2(\overline{T})\, (1 -\cos{\theta_0})\, \delta_r \right].
\end{align}
Once $T<\Tl$, the effect of temperature fluctuations vanishes and the usual expression is recovered for the perturbed axion energy density~\cite{Hwang:2009js, Marsh:2015xka}. Finally, using the axion energy density in Eq.~\eqref{eq:rho_a}, the axion overdensity becomes
\begin{equation}
    \delta_a = \frac{\dot{\theta}_0 \delta\dot{\theta}_k - \Phi \dot{\theta}_0^2 
    + \ma^2 \delta\theta_k \sin\theta_0 
    - b\, \ma^2 \delta_r (1 - \cos\theta_0)/2}{\dot\theta_0^2/2 + \ma^2\,(1-\cos\theta_0)} \,,
    \label{eq:axion_overdensity_Gen}
\end{equation}
for $T > \Tl$; the opposite case $T < \Tl$, the last term in the numerator should be discarded.

\section{Jeans threshold scale} \label{sec:jeans}
Although the axion mass is temperature dependent, radiation perturbations source axion perturbations through the explicit dependence $m_a = m_a(T)$. In this regime, the usual constant-mass Jeans argument does not directly apply: the additional radiation-induced source can continue to enhance small-scale axion perturbations after horizon entry. Once the mass saturates at $R=R_\Lambda$, however, this source is shut off. The subsequent evolution is then governed by the competition between gravitational forcing and the effective pressure of the oscillating axion field. We now estimate which modes become pressure suppressed after $R_\Lambda$.

For $R_\Lambda \lesssim R < \RRH$, we are allowed to use fluid treatment and write the equation that governs the dynamics of axions with constant mass~\cite{Arvanitaki:2009fg, Blinov:2019jqc}
\begin{equation}
    \ddot \delta_a + 2\, H\, \dot \delta_a + \omega_k^2\, \delta_a \simeq -\frac{k^2}{R^2}\, \Phi \,,
    \label{eq:axion_jeans}
\end{equation}
where we have neglected the derivatives of $\Phi$ and introduced the effective frequency $\omega_k^2 \equiv c_s^2\,k^2/R^2$, with $c_s$ the effective scale-dependent axion sound speed~\cite{Marsh:2015xka}
\begin{equation}
    c_s^2 = \frac{k^2}{k^2 + 4\, \maz^2\, R^2} \simeq \frac{k^2}{4\, \maz^2\, R^2}\,,
    \label{non_ad_axi}
\end{equation}
where the last approximation is valid for $k \ll 2\, \maz\, R$.

Equation~\eqref{eq:axion_jeans} describes a forced harmonic oscillator. The dynamics is then set by the competition between the response time of the axion pressure, $\omega_k^{-1}$, and the Hubble time $H^{-1}$, which controls the evolution of the gravitational source. When $\omega_k \ll H$, the pressure cannot react efficiently in a Hubble time and $\delta_a$ is effectively dragged by the source and continues to grow. For $\omega_k \gg H$, instead, the pressure dominates and the subsequent growth of $\delta_a$ is suppressed, leading to oscillations. The transition between both regimes occurs when $\omega_k(R_\Lambda) \sim H(R_\Lambda)$, which defines the Jeans scale
\begin{equation}
    k_J \sim R_\Lambda\, \sqrt{\maz\, H(R_\Lambda)}\,.
\end{equation}
During the EMD era, $H(R_\Lambda)=H_{\rm RH}\, x_{\Lambda}^{-3/2}$, therefore, the Jeans threshold can be found to be 
\begin{equation}
    \kappa_J \equiv \frac{k_J}{\kRH} \simeq  \sqrt{\frac{\maz}{H_{\rm RH}}}\, x_\Lambda^{1/4}\,,
\end{equation}
in good agreement with our numerical results. Modes with $\kappa > \kappa_J$ are pressure-dominated at $x_\Lambda$, stop growing, and start to oscillate at that moment. In contrast, modes with $\kappa < \kappa_J$ are mass-dominated and continue the growth accumulated during the previous stage. 

\section{Non-relativistic evolution to matter--radiation equality}
\label{sec:WKB}
For the numerical integration of the axion equations of motion during the EMD era and the subsequent evolution, we solve the coupled background and perturbation equations with a Python code based on an adaptive Runge--Kutta 45 integrator.\footnote{The code is currently in preparation for public release.} The axion evolution is computed using the numerical solution of the EMD background and perturbation system as input. Once the axion mass becomes much larger than the Hubble rate, $\ma \gg H$, the field undergoes rapid coherent oscillations, requiring increasingly small integration steps. In practice, well after reheating, around $x\sim\mathcal{O}(10^2)$, the direct Klein--Gordon integration becomes inefficient and can develop spurious oscillatory numerical artifacts.

In Section~\ref{sec:mini}, we require the evolution of axion perturbations up to MRE. The corresponding scale factor follows from entropy conservation,
\begin{equation}
x_{\rm eq}\equiv\frac{R_{\rm eq}}{\RRH}=\frac{\TRH}{T_{\rm eq}}\left(\frac{\gss(\TRH)}{\gss(T_{\rm eq})}\right)^{1/3}\simeq3.56\times10^{8}\left(\frac{\TRH}{150~{\rm MeV}}\right)\left(\frac{\gss(\TRH)}{25.91}\right)^{1/3},
\label{eq:x_eq}
\end{equation}
where we used $T_{\rm eq}=0.79~{\rm eV}$ and $\gss(T_{\rm eq})\simeq3.91$~\cite{Planck:2018vyg}. Since $x_{\rm eq}$ is many orders of magnitude larger than the range accessible to direct relativistic integration, we switch to a non-relativistic WKB description once the axion field is deep in the oscillatory regime.

We decompose both the homogeneous axion field and its perturbations into slowly varying complex amplitudes. For the homogeneous mode, we write
\begin{equation}
\theta_0(x)=\frac{1}{x\sqrt{2M(x)}}\left[\psi_0(x)e^{-i\int^x M(x')\,dx'}+\psi_0^*(x)e^{+i\int^x M(x')\,dx'}\right],
\label{eq:wkb_theta0}
\end{equation}
and similarly for the perturbation,
\begin{equation}
\delta\theta_\kappa(x)=\frac{1}{x\sqrt{2M(x)}}\left[\psi_\kappa(x)e^{-i\int^x M(x')\,dx'}+\psi_\kappa^*(x)e^{+i\int^x M(x')\,dx'}\right].
\label{eq:wkb_deltatheta}
\end{equation}
Here
\begin{equation}
M(x)\equiv\frac{\ma}{H(x)x}.
\label{eq:M_def}
\end{equation}
The matching point $x_i$ is chosen after reheating and after the axion mass has saturated to its zero-temperature value, so that $x_i\gg1$, $T(x_i)<\Tl$, and $\ma=\maz$.

During radiation domination, $H\simeq H_{\rm RH}x^{-2}$, and therefore
\begin{equation}
M(x)=\mu x,\qquad\mu\equiv\frac{\maz}{H_{\rm RH}}.
\end{equation}
The rapidly oscillating phase is then
\begin{equation}
\int^x M(x')\,dx'=\frac12\,\mu x^2+{\rm const.}
\label{eq:wkb_phase}
\end{equation}
The irrelevant constant phase is fixed by the matching convention at $x_i$.

At the matching point, the non-relativistic amplitude $\psi_\kappa(x_i)$ can be extracted from the direct numerical solution through
\begin{equation}
\psi_\kappa(x_i)\simeq e^{i\mathcal I_i}\,x_i\sqrt{\frac{M_i}{2}}\left[\delta\theta_\kappa(x_i)+\frac{i}{M_i}\left(\delta\theta_\kappa'(x_i)+\frac{1}{x_i}\delta\theta_\kappa(x_i)+\frac{M_i'}{2M_i}\delta\theta_\kappa(x_i)\right)\right],
\label{eq:psi_matching}
\end{equation}
where
\begin{equation}
\mathcal I_i\equiv\int^{x_i} M(x')\,dx',\qquad M_i\equiv M(x_i).
\end{equation}
An analogous expression is used to obtain $\psi_0(x_i)$ from $\theta_0(x_i)$ and $\theta_0'(x_i)$. In the absence of perturbative sources, $\psi_0$ is conserved at leading order in the non-relativistic expansion.

Under the transformation in Eqs.~\eqref{eq:wkb_theta0} and~\eqref{eq:wkb_deltatheta}, the perturbation equation reduces to a forced Schrödinger-like equation. Keeping the leading non-relativistic terms, one obtains
\begin{equation}
i\psi_\kappa'=\frac{\kappa^2}{2M(x)}\psi_\kappa+J_\kappa(x).
\label{eq:schrodinger_psi}
\end{equation}
During radiation domination, this becomes
\begin{equation}
i\psi_\kappa'=\frac{\kappa^2}{2\mu x}\psi_\kappa+J_\kappa(x).
\end{equation}
Since the matching point is chosen after the axion mass has become constant, the temperature-induced source satisfies $S_3=0$. For subhorizon modes, the contribution from $S_1$ is also subdominant compared with the gravitational source $S_2$. The leading source term is therefore
\begin{equation}
J_\kappa(x)\simeq\mu x\,\Phi_\kappa(x)\,\psi_0(x).
\label{eq:J_source}
\end{equation}
The solution of Eq.~\eqref{eq:schrodinger_psi} can be written in integral form. Defining the free non-relativistic phase
\begin{equation}
P_\kappa(x)\equiv\int_{x_i}^{x}\frac{\kappa^2}{2M(x')}\,dx'=\frac{\kappa^2}{2\mu}\ln\left(\frac{x}{x_i}\right),
\label{eq:Pkappa}
\end{equation}
we find
\begin{equation}
\psi_\kappa(x)\simeq e^{-iP_\kappa(x)}\left[\psi_\kappa(x_i)-i\int_{x_i}^{x}dx'\,e^{iP_\kappa(x')}J_\kappa(x')\right].
\label{eq:psi_integral_solution}
\end{equation}

To evaluate this expression up to MRE, we use the analytical form of the gravitational potential for subhorizon modes during radiation domination,
\begin{equation}
    \Phi_\kappa(x)\simeq\frac{1}{(\kappa x)^2}\left[C_1\cos\left(\frac{\kappa x}{\sqrt{3}}\right)+C_2\sin\left(\frac{\kappa x}{\sqrt{3}}\right)\right].
    \label{eq:Phi_RD_analytic}
\end{equation}
The constants $C_1$ and $C_2$ are fixed by matching $\Phi_\kappa$ and $\Phi_\kappa'$ to the numerical solution at $x_i$. This approximation captures the leading subhorizon behavior of the gravitational potential in a radiation-dominated Universe.

Finally, the axion density contrast is obtained from the interference between the homogeneous and perturbed non-relativistic fields. Averaging over the rapid oscillations, one finds
\begin{equation}
\delta_a(x)=\frac{\delta\rho_a}{\rho_a}\simeq2\,\mathrm{Re}\left(\frac{\psi_0^*(x)\psi_\kappa(x)}{|\psi_0(x)|^2}\right),
\label{eq:deltaa_NR}
\end{equation}
up to corrections suppressed by $H/\maz$ and $k^2/(\maz^2R^2)$.

\section{Isocurvature bound} \label{sec:iso}
In the pre-inflationary PQ breaking scenario considered here, the axion field is present as a light spectator during inflation and acquires quantum fluctuations of amplitude $\delta a = H_I/(2\pi)$, where $H_I$ is the inflationary Hubble scale. These fluctuations result in perturbations of the misalignment angle $\delta\theta = H_I/(2\pi f_a)$,  which generate a DM isocurvature mode uncorrelated with the primordial adiabatic curvature perturbation. In the small-angle regime, where the axion abundance scales as $\rho_a \propto \thini^2$, the corresponding isocurvature power spectrum is \cite{Hertzberg:2008wr}
\begin{equation}
    \mathcal{P}_{\mathcal{S}} = \left(\frac{\Omega_a}{\Omega_{\rm DM}}\right)^2 \left(\frac{H_I}{\pi\, f_a\, \thini}\right)^2 .
\end{equation}
Planck constrains the isocurvature fraction at the pivot scale $k_p = 0.05~\text{Mpc}^{-1}$ to $\beta_{\rm iso} \equiv \mathcal{P}_{\mathcal{S}}/(\mathcal{P}_{\mathcal{R}} + \mathcal{P}_{\mathcal{S}}) \lesssim 0.038$ at $95\%$~CL~\cite{Planck:2018vyg}, with $\mathcal{P}_{\mathcal{R}} = \mathcal{A}_s \simeq 2.101 \times 10^{-9}$. Since we want the axion to saturate the DM abundance $\Omega_a = \Omega_{\rm DM}$, one has
\begin{equation}
    H_I \lesssim \pi\, f_a\, \thini \sqrt{\frac{\beta_{\rm iso}}{1 - \beta_{\rm iso}}\, \mathcal{A}_s} \simeq 2.8 \times 10^{-5} \, f_a \, \thini .
\end{equation}
For the QCD axion benchmark of Section~\ref{sec:qcd_axion}, this bound should be evaluated using the corresponding value of $f_a\, \thini$. For example, if $f_a \simeq 10^{13}$~GeV and $\thini \simeq 1$, one obtains $H_I \lesssim 2.8 \times 10^8$~GeV. The bound is straightforwardly satisfied for sufficiently low-scale inflation. In the perturbation analysis above, we set the primordial axion isocurvature mode to zero to isolate the axion perturbations sourced by the adiabatic radiation and metric perturbations.

\bibliography{biblio_fixed}
\bibliographystyle{utphys}
\end{document}